\documentclass[%
superscriptaddress,
 amsmath,amssymb,
 aps,
 pra,
floatfix,
twocolumn
]{revtex4-2}
\usepackage[T1]{fontenc}
\usepackage[final]{graphicx}
\usepackage{dcolumn}% Align table columns on decimal point
\usepackage{amsfonts}
\usepackage{amssymb}
\usepackage{pbox}
\usepackage{amsmath}
\usepackage{amsthm}
\usepackage{mathtools}
\usepackage[normalem]{ulem}

\DeclareMathOperator{\Tr}{Tr}
\DeclareMathOperator{\diag}{diag}

\DeclareMathOperator{\re}{Re}
\DeclareMathOperator{\im}{Im}

\usepackage{caption}
\usepackage{subcaption}
\usepackage{mathtools}
\usepackage{braket}
\renewcommand\bra[1]{{\langle{#1}|}}
\renewcommand\ket[1]{{|{#1}\rangle}}

\usepackage{chngcntr}
\usepackage{multirow}
\usepackage{tabularx}
\usepackage[labelsep=period]{caption}
\date{}
\usepackage{graphicx}
\usepackage{enumerate}
\usepackage{calc}
\usepackage{indentfirst}
\usepackage{booktabs}
\usepackage{dsfont}
\usepackage{arydshln}
\usepackage{bm}
\usepackage{amsthm}
\usepackage{commath}
\usepackage[dvipsnames]{xcolor}

\newtheorem{theorem}{Theorem}
\newtheorem{proposition}[theorem]{Proposition}

\theoremstyle{definition}
\newtheorem{example}{Example}

\usepackage{natbib}
\setcitestyle{square,numbers}
\usepackage{changepage}
\usepackage[colorlinks]{hyperref}

\begin{document}

\author{Tomasz Linowski}
\affiliation{International Centre for Theory of Quantum Technologies, University of Gdansk, 80-308 Gda{\'n}sk, Poland}
\email[Corresponding author: ]{t.linowski95@gmail.com}

\author{{\L}ukasz Rudnicki}
\affiliation{International Centre for Theory of Quantum Technologies, University of Gdansk, 80-308 Gda{\'n}sk, Poland}
\affiliation{Center for Theoretical Physics, Polish Academy of Sciences, 02-668 Warszawa, Poland}

\author{Clemens Gneiting}
\affiliation{Theoretical Quantum Physics Laboratory, Cluster for Pioneering Research, RIKEN, Wako, Saitama 351-0198, Japan}
\affiliation{Center for Quantum Computing, RIKEN, Wako, Saitama 351-0198, Japan}

\title{Spectral stabilizability}

\date{\today}

\begin{abstract}
Decoherence represents a major obstacle towards realizing reliable quantum technologies. Identifying states that can be uphold against decoherence by purely coherent means, i.e., {\it stabilizable states}, for which the dissipation-induced decay can be completely compensated by suitable control Hamiltonians, can help to optimize the exploitation of fragile quantum resources and to understand the ultimate limits of coherent control for this purpose. In this work, we develop conditions for stabilizability based on the target state's eigendecomposition, both for general density operators and for the covariance matrix parameterization of Gaussian states. Unlike previous conditions for stabilizability, these spectral conditions are both necessary and sufficient and are typically easier to use, extending their scope of applicability. To demonstrate its viability, we use the spectral approach to derive upper bounds on stabilizability for a number of exemplary open system scenarios, including stabilization of generalized GHZ and W states in the presence of local dissipation and stabilization of squeezed thermal states under collective damping.
\end{abstract}

\maketitle

Uncontrolled influences of an environment generically deteriorate quantum resources, and carefully prepared quantum states rapidly lose their desired features, such as coherence and entanglement. As a consequence, this {\it decoherence} \cite{decoherence_theory_Zurek_1991,decoherence_theory_Schlosshauer_2005} renders these states less useful for practical tasks such as quantum information processing \cite{decoherence_quantum_computing_Preskill_1998,decoherence_quantum_computing_Lidar_1998,decoherence_Lidar_2014}. However, while the influence of the environment is inevitable, it can be manipulated or counteracted. 

A generic and widely applicable class of open quantum systems can be described by memoryless, i.e., Markovian, environments. The evolution of the system can then be modeled by the Gorini-Kossakowski-Lindblad-Sudarshan (GKLS) equation (also known as Lindblad equation) \cite{GKS_original,lindblad_original}, which is governed by two objects: the Hamiltonian, which describes a coherent (unitary) evolution similar to closed systems, and the dissipator, which encodes the (incoherent) decoherence effects induced by the environment. Excluding dissipative engineering, where one assumes (partial) control over the environment and hence the dissipator \cite{using_dissipation_1, Kraus2008preparation, using_dissipation_2, Lyapunov_stationarity}, manipulating the Hamiltonian remains the only alternative to counteract decoherence \cite{using_hamiltonian_1,using_hamiltonian_2}. Given a desired target state, the goal is then to identify a Hamiltonian such that the time derivative of the initial state vanishes. For a fixed dissipator, this is possible only for a specific set of states, denoted \emph{stabilizable} states \cite{stabilizability_geometric}.

The framework of stabilizability comes equipped with a set of geometric conditions \cite{stabilizability_geometric,stabilizability_cv_systems} that allow one to test whether a given state is stabilizable without the, often hard if not impossible, necessity to identify the respective control Hamiltonian. However, these geometric conditions, too, have features that render them sometimes impractical. First, they are, in general, only necessary, implying that, while they can be used to disprove a state's stabilizability, they cannot prove it. Second, except for low-dimensional systems (e.g. a single qubit or a qutrit), they assume the form of high-order polynomials in the density operator, which can make them challenging to solve even numerically \cite{stabilizing_entanglement_in_two_mode_Gaussian_states}.

In this work, we return to the first principles of stabilizability and exploit that the unitary evolution generated by the Hamiltonian can counteract the dissipation only if the dissipator leaves the initial state's eigenvalues unchanged. Based on the state's eigendecomposition, we then derive an alternative set of {\it spectral} conditions for stabilizability. These conditions are both necessary and sufficient, and they are only linear in the state's eigenvalues, regardless of the dimension of the state space. As we show, these properties distinguish the spectral conditions especially for the generic task of stabilizing a (pure) target state component against noise in an overall (unavoidably) mixed stabilizable state.

Our results apply both to the general framework of stabilizability \cite{stabilizability_geometric} and to its extension to the covariance matrix \cite{stabilizability_cv_systems}, which is typically used in studies of Gaussian continuous variables (CV) systems. In the latter case, we extend the stabilizability framework to the recently discussed \cite{quantum_gaussian_evolution_linowski_2022} evolution stemming from unitary Lindblad operators, which describes, e.g., scattering phenomena. We demonstrate the viability of the spectral stabilizability conditions with a number of explicit examples.

This article is organized as follows. In Section \ref{sec:stabilizability}, we briefly recapitulate the general concept of stabilizability. In Section \ref{sec:main_results}, we  re-derive stabilizability in terms of the target state's spectrum. The spectral conditions are then used by us in Section \ref{sec:GHZ_W} to derive fundamental limits on stabilizability of (resourceful) pure consituents in noisy mixed systems, with explicit demonstration using the $N$-qubit W and GHZ states. In Section \ref{sec:covariance_matrix}, we introduce the covariance matrix and discuss the respective geometric stabilizability conditions. The spectral approach is then extended to the covariance matrix in Section \ref{sec:covariance_matrix_stabilizing}. We conclude in Section \ref{sec:conclusion}.

\section{Stabilizability} 
\label{sec:stabilizability}
We begin with a brief summary of the stabilizability framework. Let us consider an arbitrary state $\hat{\rho}$ evolving under the GKLS (Lindblad) equation \cite{GKS_original,lindblad_original,open_systems_book_Rivas_2012}:
\begin{align} \label{eq:GKLS}
    \frac{d\hat{\rho}}{dt} = 
	    -\frac{i}{\hbar} \big[\hat{H},\hat{\rho}\big] + \gamma D(\hat{\rho}),
\end{align}
where $t$ denotes time, $\hat{H}$ is the system Hamiltonian, which is responsible for the unitary evolution, and ${D}$ is the dissipator, which encodes the effects of the interaction with the environment. Finally, $\gamma$ denotes the dissipation rate (which we introduce to keep the dissipator dimensionless). The dissipator has the general form
\begin{align} \label{eq:dissipator}
    D(\hat{\rho})=\sum_{j}\left(\hat{L}_j\hat{\rho}\hat{L}_j^\dag
        -\frac{1}{2}\big\{\hat{L}_j^\dag\hat{L}_j,\hat{\rho}\big\}\right),
\end{align}
where $\hat{L}_j$ are the Lindblad operators.

We say that a state is \emph{stabilizable} with respect to a given dissipator if there exists a Hamiltonian that is capable to counteract the effects of the dissipation. In other words, the state is stabilizable with respect to the dissipator $D$, if, for this particular dissipator and state, one can find a Hamiltonian $\hat{H}$ such that the state becomes stationary:
\begin{align} \label{eq:startionarity_condition}
    -\frac{i}{\hbar} \big[\hat{H},\hat{\rho}\big] + \gamma D(\hat{\rho}) = 0.
\end{align}
Crucially, one can often determine whether such a Hamiltonian exists without having to actually know it. In \cite{stabilizability_geometric}, the following necessary conditions for stabilizability of a state $\hat{\rho}$ were derived:
\begin{align} \label{eq:stabilizability_conditions_original}
    0=\Tr \big[ \hat{\rho}^{k}{D}(\hat{\rho}) \big]
        \textnormal{  for all  }k\in\{1,\ldots,d-1\},
\end{align}
where $d$ denotes the dimension of the Hilbert space. For states that do not exihibit degenerate eigenvalues, these conditions, which we call \textit{geometric} due to their independence of one's choice of basis, are also sufficient. As one can observe, the conditions are also independent of the Hamiltonian. Still, given a stabilizable state $\hat{\rho}$, one can recover the stabilizing Hamiltonian as \cite{stabilizability_geometric}
\begin{align} \label{eq:stabilizability_hamiltonian}
    \hat{H}=i\hbar\gamma\sum_{\substack{l,l'=0\\\lambda_l\neq\lambda_{l'}}}^{d-1}
	    \frac{\bra{\psi_l}{D}(\hat{\rho})\ket{\psi_{l'}}}
	    {\lambda_l-\lambda_{l'}}\ket{\psi_l}\bra{\psi_{l'}},
\end{align}
where $\{\lambda_l, \ket{\psi_l}\}$ is the state's eigendecomposition.

The major advantage of the stabilizability framework lies in the fact that, for a fixed dissipator, one can optimize over the set of stabilizable states to determine the most resourceful ones with respect to a given task. For example, due to its usefulness in, e.g., quantum computation  \cite{quantum_information_book,quantum_algorithms}, stabilizing entanglement was considered in \cite{stabilizability_geometric,stabilizing_entanglement_in_two_mode_Gaussian_states}. 

Irrespectively of the usefulness of the concept of stabilizability, application of the geometric conditions (\ref{eq:stabilizability_conditions_original}) is restricted, mainly for two reasons. To start with, the conditions are in general not sufficient, meaning that there can exist states which satisfy the conditions while not being stabilizable. Furthermore, the conditions are often not suitable for practical computations, as for $d$-dimensional systems they take the form of polynomial equations of up to $d$-th degree in the density operator.

In the following, we explain how both these obstacles can be overcome by approaching stabilizability from a spectral perspective.

\section{Spectral approach to stabilizability}
\label{sec:main_results}
To circumvent the problems of the geometric approach to stabilizability, we return to the main idea behind the framework and reinterpret it in the following way: since the Hamiltonian can only generate a unitary evolution, which can never alter the state's eigenvalues, a necessary condition for the state to be stabilizable is for the time evolution induced by the dissipator to leave its initial eigenvalues invariant \cite{stabilizability_geometric,stabilizability_cv_systems}. 

In other words, a state $\hat{\rho}$ may be stabilizable only if the GKLS evolution in the absence of the Hamiltonian
\begin{align} \label{eq:dissipator_evolution}
    \frac{d\hat{\rho}(t)}{dt} = \gamma{D}[\hat{\rho}(t)], \quad \hat{\rho}(0)=\hat{\rho}
\end{align}
has a solution of the form
\begin{align} \label{eq:dissipator_solution}
    \hat{\rho}(t)=\sum_{i=0}^{d-1}\lambda_i(t) \ket{\psi_i(t)}\bra{\psi_i(t)},
        \quad \frac{d\lambda_i(t)}{dt}\bigg\rvert_{t=0}=0.
\end{align}
The remaining drift of the pure constituents $\ket{\psi_i(t)}\bra{\psi_i(t)}$ at $t=0$ may, at least in principle, be counteracted by adding an appropriate Hamiltonian to the equation, yielding a truly stationary state. We will now show that this single necessary assumption for stabilizability is also sufficient for it, and thus equivalent to it. At the same time, we will derive a new set of spectral conditions for stabilizability.

We begin by observing that eq. (\ref{eq:dissipator_evolution}) must be valid in any orthonormal basis, including the one given by the eigendecomposition of the time-evolved state: 
\begin{align}
    \bra{\psi_i(t)}\frac{d\hat{\rho}(t)}{dt}\ket{\psi_{j}(t)} 
        = \gamma\bra{\psi_i(t)}{D}[\hat{\rho}(t)]\ket{\psi_{j}(t)},
\end{align}
where $i,j\in\{0,\ldots,d-1\}$. Evaluating this at $t=0$ and using eq. (\ref{eq:dissipator_solution}), we get
\begin{align}
\begin{split}
    \lambda_j \bra{\psi_i}\frac{d\ket{\psi_j}}{dt}
        + \lambda_i \frac{d\bra{\psi_i}}{dt}\ket{\psi_j}
         = \gamma\bra{\psi_i}{D}(\hat{\rho})\ket{\psi_{j}}.
\end{split}
\end{align}
Here and throughout the rest of the derivation we omit writing the time dependence explicitly -- it is assumed that all the quantities are evaluated at $t=0$. 

Let us consider what happens in the particular case where $\ket{\psi_i}$ and $\ket{\psi_j}$ correspond to the same eigenvalue $\lambda_i=\lambda_j$. Then, the above equation reduces to
\begin{align}
\begin{split}
    \lambda_j \frac{d}{dt}\left(\braket{\psi_i|\psi_j}\right)
        = \gamma\bra{\psi_i}{D}(\hat{\rho})\ket{\psi_{j}}.
\end{split}
\end{align}
The l.h.s. vanishes due to the orthonormality of the basis. Therefore, if the state is to be stabilizable, we must necessarily have
\begin{align} \label{eq:stabilizability_spectral_general_almost}
    0=\bra{\psi_i}D(\hat{\rho})\ket{\psi_j}
\end{align}
for all $i$, $j$ such that $\lambda_i=\lambda_j$.

However, crucially, it turns out to be also a sufficient condition for stabilizability: a straightforward calculation shows that, provided eq. (\ref{eq:stabilizability_spectral_general_almost}) is fulfilled, eq. (\ref{eq:startionarity_condition}) holds with the input Hamiltonian (\ref{eq:stabilizability_hamiltonian}), i.e. this Hamiltonian stabilizes the state: $\hat{\rho}(t)=\hat{\rho}(0)=\hat{\rho}$. Consequently, eq. (\ref{eq:stabilizability_spectral_general_almost}) is equivalent to stabilizability of the state $\hat{\rho}$.

%Before we state this main result more formally, \mdf{let us remark that it can be further simplified by noticing that the eigenstates $\ket{\psi_{i}}$ are not uniquely defined in degenerate subspaces of $\hat{\rho}$, where $\lambda_i=\lambda_j \textnormal{ for } i \neq j$. This gives us the freedom to choose basis states $\ket{\tilde{\psi}_{i}}$ that diagonalize the dissipator ${D}(\hat{\rho})$ in the degenerate subspaces (note that this is always possible, since ${D}(\hat{\rho})$ is a Hermitian operator), that is, ${D}(\hat{\rho}) \ket{\tilde{\psi}_{i}} \propto \ket{\tilde{\psi}_{i}}$.} The off-diagonal conditions (corresponding to $i \neq j$) are \mdf{then by definition satisfied, and we are left with the diagonal ($i = j$) conditions, $0=\bra{\tilde{\psi}_i}D(\hat{\rho})\ket{\tilde{\psi}_i}$ for $i \in \{0, \dots, d-1\}$. This shows that the number of stabilizability conditions matches the dimension of the Hilbert space. In some practical situations, where diagonalizing the dissipator ${D}(\hat{\rho})$ is challenging, it may still be preferable to work with the conditions~(\ref{eq:stabilizability_spectral_general_almost}).}

We summarize this in the form of a proposition.

\begin{proposition}[Spectral conditions for stabilizability] 
\label{th:stabilizability_spectral_general}
Let $\{\lambda_j, \ket{\psi_j}\}$ be the eigendecomposition of the state $\hat{\rho}$. The state $\hat{\rho}$ is stabilizable with respect to the dissipator $D$ if and only if
\begin{align} \label{eq:stabilizability_spectral_general}
\begin{split}
    0=\bra{\psi_i}D(\hat{\rho})\ket{\psi_j} \textnormal{ for all } i,j \textnormal{ such that } \lambda_i=\lambda_j.
\end{split}
\end{align}
%\begin{subequations} \label{eq:stabilizability_spectral_general}
%\begin{align}
%\begin{split}
 %   \mdf{0=\bra{\psi_i}D(\hat{\rho})\ket{\psi_j} \textnormal{ for all } i,j \textnormal{ such that } \lambda_i=\lambda_j}.
%\end{split}
%\end{align}
%\mdf{If the basis states $\ket{\psi_i}$ in addition diagonalize the Hermitian operator $D(\hat{\rho})$ in degenerate subspaces of $\hat{\rho}$, then the state $\hat{\rho}$ is stabilizable if and only if}
%\begin{align}
%\begin{split}
 %   \mdf{0=\bra{\psi_i}D(\hat{\rho})\ket{\psi_i} \textnormal{ for all } i \in \{0, \dots, d-1\}}.
%\end{split}
%\end{align}
%\end{subequations}
\end{proposition}

Let us discuss this result.

Comparing the spectral conditions to their geometric counterpart (\ref{eq:stabilizability_conditions_original}), we find that the former assume knowledge of the eigendecomposition of the state. While this may represent a difficulty if we want to test the stabilizability of a given state, we observe that, from a practical point of view, the geometric approach shares this obstacle in some ways. On the one hand, we need to know the spectrum of the state in order to decide if the geometric conditions are not only necessary but also sufficient. On the other hand, the state's eigendecomposition is required in order to determine the counteracting control Hamiltonian according to~(\ref{eq:stabilizability_hamiltonian}).

Nonetheless, in principle, the spectral conditions can be solved without having to know the eigendecomposition of the state: it is straightforward to see that eq. (\ref{eq:stabilizability_spectral_general}) is equivalent to \footnote{Necessity can be easily shown by multiplying eq. (\ref{eq:startionarity_condition}) by $\hat{X}$ and taking the trace. Then, the requirement $[\hat{X}, \hat{\rho}]=0$ makes the Hamiltonian part vanish. Sufficiency follows by direct inspection.}
\begin{align} \label{eq:commutator_approach}
\begin{split}
    0 = \Tr D(\hat{\rho}) \hat{X} \textnormal{ for all } \hat{X} \textnormal{ such that } [\hat{X}, \hat{\rho}] = 0.
\end{split}
\end{align}
The set of all such $\hat{X}$ can always be easily found even for large $d$, since $[\hat{X},\hat{\rho}]=0$ is a linear equation for the matrix elements of $\hat{X}$. While this offers an interesting and potentially useful reformulation, solving the actual conditions (\ref{eq:commutator_approach}) is typically not easier than finding the eigendecomposition of the state in the first place. Moreover, such ``commutator'' approach lacks the immediate physical interpretation of the spectral perspective, which is why in the following we focus on the latter.

We remark that the spectral stabilizability conditions can be refined further. To this end, we observe that eq. (\ref{eq:stabilizability_spectral_general}) is equivalent to the vanishing of the dissipator $D(\hat{\rho})$ on all the eigenspaces of the target state. In particular, this means that in the degenerate subspaces (but only there), where the state's eigendecomposition is not uniquely defined, we are free to choose eigenstates $\ket{\tilde{\psi}_j}$ such that ${D}(\hat{\rho}) \ket{\tilde{\psi}_{j}} \propto \ket{\tilde{\psi}_{j}}$ [this is always possible, since ${D}(\hat{\rho})$ is a Hermitian operator]. The off-diagonal conditions (corresponding to $i \neq j$) are then by definition satisfied, and we are left only with the diagonal ($i = j$) ones:
\begin{align}
\begin{split}
    0=\bra{\tilde{\psi}_j}D(\hat{\rho})\ket{\tilde{\psi}_j} 
        \textnormal{ for all } j \in \{0, \dots, d-1\}.
\end{split}
\end{align}
This shows that the number of stabilizability conditions formally matches the dimension of the Hilbert space. In practial applications, however, it may be preferable to avoid an additional diagonalization step and to resort to the agnostic conditions (\ref{eq:stabilizability_spectral_general}).

The spectral approach possesses three potential advantages compared to the original, geometric one. Firstly, while the original conditions are in general only necessary for stabilizability, the spectral conditions are both necessary and sufficient. Secondly, they are only linear in the state's eigenvalues, which makes the generic problem of finding stabilizable states diagonal in a given basis particularly easy. Thirdly, as we will elaborate in the examples, the spectral approach allows us to directly address the stabilizability of desired, resourceful target state components in an overall mixed state.

We note that it is straightforward to show that for non-degenerate states, the spectral conditions (\ref{eq:stabilizability_spectral_general}) are equivalent to
\begin{align} \label{eq:orthogonal_fluxes}
    0=\vec{\lambda}\cdot\vec{F}_k\textnormal{  for all  }k\in\{1,\ldots,d-1\},
\end{align}
where the $j$-th component of each of the vectors $\vec{F}_k$ is defined as
\begin{align}
    (F_k)_j\coloneqq
        \Tr[\ket{\psi_k}\bra{\psi_k}{D}(\ket{\psi_j}\bra{\psi_j})].
\end{align}
Suppose we are interested in stabilizable states diagonal in some basis $\ket{\psi_j}$ (we will discuss several examples for this case below). Eq. (\ref{eq:orthogonal_fluxes}) implies that the eigenvalues of all such states can be found as the vectors $\vec{\lambda}$ orthogonal to the ``dissipative fluxes'' $\vec{F}_k$. This again reinforces the idea that eigenvalues of a stabilizable state must be invariant under the action of the dissipator.

Let us remark that eq. (\ref{eq:orthogonal_fluxes}) is reminiscent of, and generalizes, the stabilizability condition for a single qubit \cite{stabilizability_geometric}, which, when expressed in terms of the Bloch representation, reads 
\begin{align} \label{eq:orthogonal_fluxes_qubit}
    0=\vec{r}\cdot\vec{F}_{\textnormal{qubit}},
\end{align}
where $r_j\coloneqq \Tr\hat{\rho}\,\hat{\sigma}_j$ is the qubit's Bloch vector, $\hat{\sigma}_j$ are the Pauli matrices and the Bloch-picture dissipative flux is given by
\begin{align}
    (F_{\textnormal{qubit}})_j = \Tr[\hat{\sigma}_j D(\hat{\rho})].
\end{align}

The following examples are meant to illustrate the advantages of the spectral approach in conceptually simple settings. The application of the spectral conditions to more complex, practical problems follows in the next section.

\begin{example}[Damping of a qubit] \label{ex:qubit_damping}
The goal of our first example is to clarify whether the maximally mixed state of a two-level system (i.e., $d=2$),
\begin{align} \label{eq:qubit_damping_maximally_mixed_state}
    \hat{\rho}_{\textnormal{mix}} = \frac{1}{2}\hat{\mathds{1}}_2 ,
\end{align}
is stabilizable under amplitude damping
\begin{align}
    \hat{L} = \ket{0}\bra{1},
\end{align}
where the excited state decays into the ground state. 

In this case, we obtain
\begin{align} \label{eq:qubit_damping_disspator}
    D(\hat{\rho}_{\textnormal{mix}}) = -\frac{1}{2} \hat{\sigma}_3,
\end{align}
with $\hat{\sigma}_3=\ket{0}\bra{0}-\ket{1}\bra{1}$ being the Pauli $z$ matrix. Consequently, the state $\hat{\rho}_{\textnormal{mix}}$ fulfills the geometric conditions (\ref{eq:stabilizability_conditions_original}) for stabilizability, which in this case reduce to just one equation:
\begin{align}
    \Tr \big[ \hat{\rho}_{\textnormal{mix}} D(\hat{\rho}_{\textnormal{mix}}) \big] = -\frac{1}{4} \Tr \hat{\sigma}_3 = 0.
\end{align}
In other words, the geometric conditions do not exclude the stabilizability of the maximally mixed state.

Let us now evaluate the corresponding spectral conditions~(\ref{eq:stabilizability_spectral_general}). To this end, we first note that the considered state~(\ref{eq:qubit_damping_maximally_mixed_state}) is degenerate, which leaves us to choose a basis. In the eigenbasis of the dissipator~(\ref{eq:qubit_damping_disspator}) it is enough to test only one of the diagonal conditions, yielding
\begin{align} \label{eq:damping}
    \bra{0}D(\hat{\rho}_{\textnormal{mix}})\ket{0} 
        = -\bra{1}D(\hat{\rho}_{\textnormal{mix}})\ket{1} = \frac{1}{2} \neq 0,
\end{align}
which unambiguously clarifies that the maximally mixed state is not stabilizable under amplitude damping. Note that there is no contradiction, since the geometric conditions are not sufficient for stabilizability of degenerate states.

It is instructive to evaluate the spectral conditions in a basis that does not diagonalize the dissipator~(\ref{eq:qubit_damping_disspator}), e.g., the eigenbasis of $\hat{\sigma}_1=\ket{+}\bra{+}-\ket{-}\bra{-}$, with $\ket{\pm}=\frac{1}{\sqrt{2}}\left(\ket{0}\pm\ket{1}\right)$. One then finds that
\begin{align}
    \bra{+}D(\hat{\rho}_{\textnormal{mix}})\ket{+} 
        = \bra{-}D(\hat{\rho}_{\textnormal{mix}})\ket{-} = 0,
\end{align}
that is, the diagonal conditions alone can in general not exclude stabilizability of the state (\ref{eq:qubit_damping_maximally_mixed_state}). Only the off-diagonal conditions
\begin{align}
    \bra{+}D(\hat{\rho}_{\textnormal{mix}})\ket{-} = \bra{-}D(\hat{\rho}_{\textnormal{mix}})\ket{+} 
        = \frac{1}{2} \neq 0
\end{align}
deliver the relevant information for this basis choice.
\end{example}

\begin{example}[Damping of diagonal states] \label{ex:simple_damping}
For our second example, we move to an infinitely-dimensional Hilbert space and consider a natural generalization of the qubit damping operator (\ref{eq:damping}) given by the annihilation operator: $\hat{L}=\hat{a}$. Physically, such dissipation may describe the spontaneous loss of particles in the system due to leakage into the environment. The model is commonly used to emulate the presence of simple noise in the system.

For simplicity, we restrict ourselves to states which are diagonal in the number basis, i.e.
\begin{align} \label{eq:rho_c}
    \hat{\rho}_c=\sum_{j=0}^{\infty}\lambda_j\ket{j}\bra{j}.
\end{align}
Such states include, e.g., thermal states of the harmonic oscillator. 

Making use of the original conditions (\ref{eq:stabilizability_conditions_original}), we find that, for the problem at hand, stabilizable states must fulfill
\begin{align} \label{eq:for_Vandermonde}
    0=\sum_{j=0}^\infty \lambda_j^k\left[(j+1)\lambda_{j+1}-j\lambda_j\right]
        \textnormal{  for all  }k\in\mathbb{N}_+.
\end{align}
Solving this infinite hierarchy of equations or proving that it has no solutions is a difficult task, unless special assumptions are made. 

For example, one can show that the equation has no solutions for finite-rank, non-degenerate states. To this end, we can rewrite eq. (\ref{eq:for_Vandermonde}) as a matrix equation 
\begin{align}
    0= \mathcal{M}\vec{v},
\end{align}
where $\mathcal{M}_{kj}\coloneqq \lambda_j^k$ and $v_j \coloneqq (j+1)\lambda_{j+1}-j\lambda_j$. By construction, $\mathcal{M}$ is a Vandermonde matrix, with $\det \mathcal{M} = \prod_{l<l'} (\lambda_l-\lambda_{l'})$ \cite{stabilizability_geometric}. Obviously, if the eigenvalues are non-degenerate, this determinant is non-zero and hence, at least in the case of finite-rank states, the equation holds only if $\vec{v}=0$, i.e., 
\begin{align} \label{eq:spectral_super_Vandermonde}
    0 = (j+1)\lambda_{j+1}-j\lambda_j
        \textnormal{  for all  }j\in\mathbb{N}.
\end{align}
This hierarchy has no nontrivial solutions. The condition for $j=0$ implies $\lambda_{1}=0$. In turn, the condition for $j=1$ implies $\lambda_{2}=0$, and so on. Therefore, the only solution is $\lambda_j=\delta_{j0}$, which corresponds to the vacuum state $\hat{\rho}_c=\ket{0}\bra{0}$.

However, if eigenvalue degeneracy is not a priori excluded, the argument with the Vandermonde matrix is not applicable. On the other hand, it is straightforward to calculate the corresponding spectral stabilizability conditions 
\begin{align}
    0 = \bra{i}D(\hat{\rho}_c)\ket{j} \textnormal{ for all } i,j \textnormal{ such that } \lambda_i=\lambda_j.
\end{align}
It is easy to show that the diagonal conditions ($i=j$) are equivalent to eq. (\ref{eq:spectral_super_Vandermonde}), while the offdiagonal conditions vanish per definition, as $D(\hat{\rho}_c)$ is diagonal in the number basis $\ket{i}$. This implies that the vacuum state is the only state in the family (\ref{eq:rho_c}) which is stabilizable with respect to damping, regardless of eigenvalue degeneracy and the rank of the state.

As seen, whereas the original necessary conditions let us characterize the family of stabilizable states only for a certain subclass of states, the spectral conditions let us characterize it in the general case (and with less effort). We remark that the case of $\hat{L}=\hat{a}^\dag$, i.e., spontaneous particle production, can be treated in a similar fashion, yielding no stabilizable states.
\end{example}

\section{Stabilizing resourceful pure constituents of noisy mixtures}
\label{sec:GHZ_W}
As discussed above, the spectral approach to stabilizability is the most useful when the target state's eigenstructure is at least partially known. This includes the important class of problems where the task is to stabilize some resourceful pure eigenstate $\ket{\mathcal{P}}$ of a generically mixed quantum state
\begin{align} \label{eq:rho_P}
    \hat{\rho}_\mathcal{P} \coloneqq p_\mathcal{P} \ket{\mathcal{P}}\bra{\mathcal{P}} 
        + (1-p_\mathcal{P}) \hat{\sigma}_\mathcal{P},
\end{align}
where $p_\mathcal{P}\in[0,1]$ and $\hat{\sigma}_\mathcal{P}$ represents an unknown ``noise'' state orthogonal to $\ket{\mathcal{P}}$. Note that, by definition, $\hat{\sigma}_\mathcal{P}\ket{\mathcal{P}}=0$, however $\hat{\sigma}_\mathcal{P}$ is otherwise unconstrained.

We now wish to answer the following question: assuming that $\hat{\rho}_\mathcal{P}$ evolves under some dissipative GKLS evolution, what is the maximum value of $p_\mathcal{P}$, for which $\hat{\rho}_\mathcal{P}$ may be stabilized?

Valuable information is obtained by considering just one of the spectral conditions (\ref{eq:stabilizability_spectral_general}), namely, the one associated solely with $\ket{\mathcal{P}}$:
\begin{align}
\begin{split}
    0 & = \bra{\mathcal{P}}D(\hat{\rho}_\mathcal{P})\ket{\mathcal{P}} \\
    & = \bra{\mathcal{P}}\left[p_\mathcal{P} D(\ket{\mathcal{P}}\bra{\mathcal{P}})\
     + (1-p_\mathcal{P}) D(\hat{\sigma}_\mathcal{P})\right]\ket{\mathcal{P}}.
\end{split}
\end{align}
Solving for $p_\mathcal{P}$, we obtain
\begin{align} \label{eq:P_condition}
\begin{split}
    p_{\mathcal{P}} = 
        \frac{\bra{\mathcal{P}}D(\hat{\sigma}_\mathcal{P})\ket{\mathcal{P}}}
        {\bra{\mathcal{P}}D(\hat{\sigma}_\mathcal{P})\ket{\mathcal{P}}
        -\bra{\mathcal{P}}D(\ket{\mathcal{P}}\bra{\mathcal{P}})\ket{\mathcal{P}}}.
\end{split}
\end{align}
As we will now show through examples of $N$-qubit maximally entangled states, the r.h.s. of the above equation can be upper-bounded by simple functions of $N$, leading to meaningful restrictions on stabilizability of the state $\ket{\mathcal{P}}$ inside the mixture (\ref{eq:rho_P}). Let us emphasize again that a similar reasoning cannot be easily applied to the geometric approach, due to its uniform treatment of the system's eigenstates.

\begin{example}[Generalized GHZ state in a damped multi-qubit system] \label{ex:GHZ}
As our first choice for $\ket{\mathcal{P}}$, we pick the generalized Greenberger–Horne–Zeilinger (GHZ) state, an $N$-qubit maximally entangled state useful, e.g., in quantum computing \cite{three_qubits_entanglement_types,entanglement_resource_theory_unique_maximal_Contreras_2019,generalized_GHZ_state_computing_2019}: 
\begin{align}
    \ket{\Phi} \coloneqq 
        \frac{1}{\sqrt{2}}\left(\ket{0_1, \ldots, 0_N}+\ket{1_1, \ldots, 1_N}\right).
\end{align}
Note that the GHZ state is typically denoted by $\ket{GHZ}$, however, we use $\ket{\Phi}$ to shorten notation. As for the dissipator, we assume local dampings:
\begin{align} \label{eq:local_damping}
    \hat{L}_j = \ket{0_j}\bra{1_j}, \quad j=1,\ldots,N.
\end{align}
Here, the index $j$ refers to the fact that $\hat{L}_j$ acts only on the $j$-th qubit, leaving the remaining qubits unaffected. Such local dissipation generically acts adversely to entanglement.

Substituting $\mathcal{P}=\Phi$ into eq. (\ref{eq:P_condition}), we obtain
\begin{align} \label{eq:GHZ_condition}
\begin{split}
    p_{\Phi} = 
        \frac{\bra{\Phi}D(\hat{\sigma}_\Phi)\ket{\Phi}}
        {\bra{\Phi}D(\hat{\sigma}_\Phi)\ket{\Phi}
        -\bra{\Phi}D(\ket{\Phi}\bra{\Phi})\ket{\Phi}}.
\end{split}
\end{align}
The r.h.s. depends only on two quantities: $\bra{\Phi}D(\hat{\sigma}_\Phi)\ket{\Phi}$ and $\bra{\Phi}D(\ket{\Phi}\bra{\Phi})\ket{\Phi}$. By direct calculation, we find that the latter equals
\begin{align} \label{eq:GHZ_term_1}
\begin{split}
    \bra{\Phi}D(\ket{\Phi}\bra{\Phi})\ket{\Phi} = -1.
\end{split}
\end{align}
Note that this immediately implies that $|\Phi\rangle$ itself is not stabilizable. To determine the former, we use the fact that, for any states $\hat{\rho}_1$, $\hat{\rho}_2$ and any dissipator,
\begin{align}
\begin{split}
    \Tr \hat{\rho}_1 D(\hat{\rho}_2) = \Tr \tilde{D}(\hat{\rho}_1) \hat{\rho}_2,
\end{split}
\end{align}
where [cf. eq. (\ref{eq:dissipator})]
\begin{align} \label{eq:dissipator_tilde}
    \tilde{D}(\hat{\rho})=\sum_{j}\left(\hat{L}_j^\dag\hat{\rho}\hat{L}_j
        -\frac{1}{2}\big\{\hat{L}_j^\dag\hat{L}_j,\hat{\rho}\big\}\right).
\end{align}
Applied to the case at hand, we obtain
\begin{align} \label{eq:GHZ_term_2}
\begin{split}
    \bra{\Phi}D(\hat{\sigma}_\Phi)\ket{\Phi} 
        = \Tr\hat{\sigma}_\Phi\tilde{D}(\ket{\Phi}\bra{\Phi}) = \Tr \hat{\sigma}_\Phi \hat{\chi},
\end{split}
\end{align}
where
\begin{align} \label{eq:chi_eigendecomposition}
\begin{split}
    \hat{\chi} = \frac{1}{N}\sum_{j=1}^N\ket{0_1,\ldots,1_j,\ldots,0_N}\bra{0_1,\ldots,1_j,\ldots,0_N}
\end{split}
\end{align}
formally describes a density operator. Substituting eqs (\ref{eq:GHZ_term_1}, \ref{eq:GHZ_term_2}) into the condition (\ref{eq:GHZ_condition}) and rearranging yields
\begin{align}
\begin{split}
    p_\Phi = \frac{\Tr \hat{\sigma}_\Phi \hat{\chi}}
    {1 + \Tr \hat{\sigma}_\Phi \hat{\chi}}.
\end{split}
\end{align}
It is easy to show that, if $\hat{\sigma}$ is a density operator and $\hat{A}$ is Hermitian, then $\Tr \hat{\sigma} \hat{A}$ can be upper bounded by the largest eigenvalue of $\hat{A}$ \footnote{Let $\hat{A}$ have the eigendecomposition $\hat{A}=\Sigma_j A_j \ket{A_j}\bra{A_j}$. Then $\Tr \hat{\sigma} \hat{A} = \Sigma_{j} A_j \Tr \hat{\sigma} \ket{A_j}\bra{A_j} \leqslant (\max_l A_l) \Tr \hat{\sigma} \Sigma_{j} \ket{A_j}\bra{A_j} = (\max_l A_l) \Tr \hat{\sigma} = \max_l A_l$.}. In the case at hand, it immediately follows from eq. (\ref{eq:chi_eigendecomposition}) that all eigenvalues of $\hat{\chi}$ are equal to $1/N$ and hence
\begin{align} \label{eq:GHZ_p}
\begin{split}
    p_\Phi \leqslant \frac{1}{N+1}.
\end{split}
\end{align}
We observe that, as $N$ grows, $p_\Phi$ approaches zero, increasingly limiting stabilizability of the GHZ state in the mixture. Note, however, that the decay of $p_\Phi$ indicated in (\ref{eq:GHZ_p}) may be algebraic, in contrast to the exponential decay $1/2^N$ of the fraction of $\Phi$ in the maximally mixed state, which can serve as a reference.

As a special case, let us consider two qubits, or $N=2$, where the generalized GHZ state reduces to the Bell state $\ket{\Phi_+}$, known, e.g., from quantum teleportation protocols \cite{Bell,quantum_information_book,quantum_teleportation},
\begin{align}
    \ket{\Phi} \xrightarrow[N \to 2]{} \ket{\Phi_+} \coloneqq \frac{1}{\sqrt{2}}\left(\ket{00}+\ket{11}\right).
\end{align}
In this case, we obtain $p_\Phi \leqslant 1/3$, and the respective fidelity with the Bell state $\ket{\Phi_+}$ is then
\begin{align}
    \mathcal{F}_{\Phi_+}(\hat{\rho}_{\Phi_+}) 
        = \bra{\Phi_+} \hat{\rho}_{\Phi_+} \ket{\Phi_+} \leqslant \frac{1}{3} .
\end{align}

In contrast, if the stabilizable fidelity with $\ket{\Phi_+}$ is optimized directly (again under local damping), it was previously found \cite{using_hamiltonian_1} that the fidelity can reach
\begin{align}
    \mathcal{F}_{\Phi_+}(\hat{\rho}_{\rm fid}) = \bra{\Phi_+}\hat{\rho}_{\rm fid}\ket{\Phi_+} = (3+\sqrt{5})/8 \approx 0.65 ,
\end{align}
where
\begin{align}
\hat{\rho}_{\rm fid} = \frac{1}{1+\varphi^2} \big(\frac{1}{4} \hat{\mathds{1}}_4 + \varphi^2 \ket{00}\bra{00} + \frac{\varphi}{2} \ket{00}\bra{11} + \frac{\varphi}{2} \ket{11}\bra{00} \big)
\end{align}
and $\varphi=(1+\sqrt{5})/2$ is the golden ratio. While $\mathcal{F}_{\Phi_+}(\hat{\rho}_{\rm fid})$ is significantly larger than $\mathcal{F}_{\Phi_+}(\hat{\rho}_\Phi)$, it is important to realize that $\ket{\Phi_+}$ is not an eigenstate of $\hat{\rho}_{\rm fid}$, and hence the higher fidelity compared to $\hat{\rho}_\Phi$ does not reflect an increased capability to harness and deploy $\hat{\rho}_{\rm fid}$ for tasks that require the Bell state $\ket{\Phi_+}$. Indeed, even the separable pure state $\ket{00}$ achieves $\mathcal{F}_{\Phi_+}(\ket{00}\bra{00}) = 1/2$, although it is entirely dysfunctional in tasks that require entanglement. This example may serve to demonstrate that the spectral stabilization developed here, quantified by the eigenvalue $p_\Phi$, provides the arguably most precise assessment of to what extent a desired target state can be stabilized in the presence of dissipation.
\end{example}

\begin{example}[Generalized W state in a damped multi-qubit system] \label{ex:W}
For comparison, let us consider a similar scenario as in the previous example, but with the GHZ state replaced by the the generalized W state:
\begin{align}
    \ket{W} \coloneqq \frac{1}{\sqrt{N}}\sum_{j=1}^N \ket{0_1,\ldots,1_j,\ldots,0_N}.
\end{align}
Similarly to the GHZ state, the W state is also maximally entangled, albeit in an inequivalent way \cite{three_qubits_entanglement_types}.

Proceeding in a complete analogy to the previous example, we obtain
\begin{align}
\begin{split}
    p_W = \frac{\Tr \hat{\sigma}_W \hat{\zeta}}
    {\frac{1}{N-1} + \Tr \hat{\sigma}_W \hat{\zeta}},
\end{split}
\end{align}
where
\begin{align} \label{eq:zeta_state}
\begin{split}
    \hat{\zeta} = \frac{1}{N}\sum_{j=1}^N\ket{\zeta_{j}}\bra{\zeta_{j}}
\end{split}
\end{align}
is formally a density operator, with
\begin{align}
\begin{split}
    \ket{\zeta_{j}} \coloneqq \frac{1}{\sqrt{N-1}}\sum_{\substack{k=1\\k\neq j}}^N
        \ket{0_1,\ldots,1_j,\ldots,1_k,\ldots,0_N}.
\end{split}
\end{align}
$\Tr \hat{\sigma}_W \hat{\zeta}$ can again be upper bounded by the largest eigenvalue of $\hat{\zeta}$. This time, however, the $\ket{\zeta_j}$ are not mutually orthogonal and therefore the eigenvalues of $\hat{\zeta}$ cannot be simply read off from eq. (\ref{eq:zeta_state}). Instead, we observe that
\begin{align}
\begin{split}
    \sqrt{\frac{2}{N(N-1)}}\sum_{j=1}^{N-1}\sum_{k=j+1}^N
        \ket{0_1,\ldots,1_j,\ldots,1_k,\ldots,0_N}
\end{split}
\end{align}
is an eigenstate of $\hat{\zeta}$ with eigenvalue $2/N$. Based on our findings for $N\leqslant 13$, we conjecture that this eigenvalue is always the largest: for $N \in \{2,3,4\}$ this is obvious, for $N \in \{ 5, \ldots, 10\}$ we checked it by an explicit calculation, while for $N \in \{ 11, 12, 13\}$ we checked it numerically \footnote{In fact, based on our findings we conjecture more generally that for all $N$ the eigenvalues of $\hat{\zeta}$ are: $2/N$ with no degeneracy, $(N - 2)/(N (N - 1))$ with degeneracy $N-1$ and zero with degeneracy $2^N-N$.}.

Assuming the conjecture holds, we have
\begin{align} \label{eq:W_p}
\begin{split}
    p_W \leqslant \frac{2N-2}{3N-2},
\end{split}
\end{align}
which approaches the positive value of $2/3$ with growing $N$. Comparing this to the analogous result (\ref{eq:GHZ_p}) for the GHZ state, we can see that the W state appears significantly more robust against noise, in agreement with previous findings \cite{W_GHZ_comparison_robustness_Carvahlo_2004}.

Once again, it is instructive to consider the two-qubit case, for which the W state reduces to another Bell state
\begin{align}
    \ket{W} \xrightarrow[N \to 2]{} \ket{\Psi_+} \coloneqq \frac{1}{\sqrt{2}}\left(\ket{01}+\ket{10}\right).
\end{align}
In this case, according to eq. (\ref{eq:W_p}) the probability $p_W$ is bounded from above by $1/2$. As it is easy to check, this value is obtained by the noise state $\hat{\sigma}_W = \ket{11}\bra{11}$. The fidelity
\begin{align}
    \mathcal{F}_{\Psi_+}(\hat{\rho}_{\Psi_+}) = \bra{\Psi_+}\hat{\rho}_{\Psi_+}\ket{\Psi_+}
\end{align}
between the corresponding two-qubit state [eq. (\ref{eq:rho_P}) with $\mathcal{P}=\Psi_+$ and $\hat{\sigma}_{\Psi_+} = \ket{11}\bra{11}$] and the Bell state $\ket{\Psi_+}$
equals $\bar{\mathcal{F}}_{\Psi_+}=1/2$, reproducing the result from \cite{stabilizability_geometric}.

Let us remark that, so far, the best known stabilizing Hamiltonians achieve at most $\bar{\mathcal{F}}_{W}=1/2$ (i.e., $p_W = 1/2$), not only for two qubits, but for all $N$ \cite{stabilizability_geometric}. We leave it to future analysis to identify Hamiltonians (if existing) that approach the potentially superior limit capacity (\ref{eq:W_p}).
\end{example}

\section{Gaussian states and the covariance matrix} 
\label{sec:covariance_matrix}
In principle, the framework of stabilizability, as defined above, can be applied to arbitrary quantum systems, including continuous-variable (CV) ones. However, in practice, especially when dealing with Gaussian states, CV systems are often more easily described not by their (typically complex) density operators, but by their low-order correlation functions, conveniently collected in the covariance matrix (see below for definitions). Consequently, in such cases, it is the covariance matrix that is stabilized. 

In this section, we briefly summarize the notion of covariance matrix, its evolution and stabilizability. The spectral approach to the latter is discussed by us in the next section.

\subsection{Symplectic picture}
Let us consider an $N$-mode Hilbert space $\mathcal{H} = \bigotimes_{k=1}^N\mathcal{H}_k$ equipped with $N$ pairs of mode quadratures $\hat{x}_k, \hat{p}_k$, as well as the vector
\begin{align} \label{eq:xi}
\hat{\vec{\xi}} \coloneqq \big(\hat{x}_1, \hat{p}_1, \ldots, \hat{x}_N, \hat{p}_N\big)^T.
\end{align}
Since the mode quadratures form a basis of operators acting in the $N$-mode Hilbert space, the state of the system can be fully described \cite{Ivan_2012} by the complete ($n=1,\ldots,\infty$) set of \emph{$n$-th order correlation functions (correlations)} of the form 
\begin{align} \label{eq:moment_definition}
\braket{\hat{\xi}_{l_1}\ldots\hat{\xi}_{l_n}}
    \coloneqq\Tr\big(\hat{\rho}\,\hat{\xi}_{l_1}\ldots\hat{\xi}_{l_n}\big),
\end{align}
which we also call \emph{$n$-th moments} for short. In many studies, especially those involving Gaussian states, i.e. states with Gaussian characteristic functions \cite{two-mode_gaussian_etc_proper_norm,cv_systems_gaussian_states,gaussian_optics}, it is enough to consider only the first and second moments. The advantage is that, in contrast to the infinitely-dimensional density operator, the first two moments are completely described by a finite number of degrees of freedom \footnote{More precisely, for an $N$-mode system, there are $2N$ first moments and $4N^2$ second moments. Taking into account the fact that some of the second moments are co-dependent (due to the canonical commutation relations) yields a total of $N(2N+3)$ independent parameters.}.

The first moments contain solely local information, and as such are irrelevant from the point of view of stabilizing most quantum resources such as, e.g., entanglement or purity of $N$-mode states. Being concerned predominantly with practical applications of stabilizability, in the remainder of this work, we assume that the first moments vanish.

The second moments are conveniently collected in the $2N\times 2N$ \emph{covariance matrix} through the mean values (\ref{eq:moment_definition}) of the quadratures' anticommutators:
\begin{align} \label{eq:covariance_matrix}
    V_{kk'}\coloneqq\frac{1}{2}
        \braket{\big\{\hat{\xi}_k,\hat{\xi}_{k'}\big\}},
\end{align}
Any valid covariance matrix has to be positive and fulfill the Heisenberg uncertainty principle:
\begin{align} \label{eq:uncertainty_principle}
\sqrt{\braket{\hat{x}_k^2}-\braket{\hat{x}_k}^2}
	\sqrt{\braket{\hat{p}_k^2}-\braket{\hat{p}_k}^2}
	\geqslant \frac{\hbar}{2},
\end{align}
where $k\in\{1,\ldots,N\}$, equivalent to \cite{two-mode_gaussian_etc_proper_norm}
\begin{align} \label{eq:Heisenberg_uncertainty_principle}
    V+\frac{i}{2}J \geqslant 0.
\end{align}
Here, $J$ is the $2N\times 2N$ \emph{symplectic form}, defined in terms of the canonical commutation relations as
\begin{align} \label{eq:symplectic_form}
    J_{kk'} \coloneqq -\frac{i}{\hbar}\big[\hat{\xi}_k,\hat{\xi}_{k'}\big]
\end{align}
and explicitly equal to
\begin{align} \label{eq:symplectic_form_explicit}
    J = \bigoplus_{k=1}^N J_2, 
        \quad J_2 \coloneqq
        \begin{bmatrix}
        0&1\\
        -1&0
        \end{bmatrix},
\end{align}
where $J_2$ is an ordinary $2\times 2$ matrix. Note that $J$ fulfills the characteristic properties 
\begin{align} \label{eq:J_properties}
    J^T=J^{-1}=-J, \quad J^2=-\mathds{1}_{2N}.
\end{align}
The symplectic form defines the symplectic group $Sp(2N,\mathbb{R})$ consisting of matrices $K$ of size $2N \times 2N$, such that \footnote{Some define symplectic matrices by an alternative relation $K^T J K = J$. However, both definitions are completely equivalent, since when $K$ is symplectic, so is $K^T$. Indeed, assume, e.g., $K J K^T = J$ holds. Then $K^T = J K^{-1} J^{-1}$, which substituted into $K^T J K$ yields $J$. Proof in the other direction is analogous.}
\begin{align} \label{eq:symplectic_matrices_definition}
\begin{split}
    KJK^T=J.
\end{split}
\end{align}

The pair $(V,\vec{\xi})$ defines the \emph{symplectic picture} (also known as the \emph{covariance matrix picture}) of quantum states. All the standard notions known from the density operator picture translate in a natural way to the symplectic picture. In particular, just like any density operator can be diagonalized by a unitary operation and is then described by its eigenvalues, any covariance matrix can be diagonalized by a symplectic operation and is then described by its symplectic eigenvalues
\begin{align} \label{eq:symplectic_eigenvalues}
\begin{split}
    1/2\leqslant \nu_1 \leqslant \ldots \leqslant \nu_N.
\end{split}
\end{align}
The symplectic eigenvalues come in pairs, i.e. the diagonalized covariance matrix reads $V_{\diag}=\diag(\nu_1,\nu_1,\ldots,\nu_N,\nu_N)$. They can be computed from the eigenvalues of the matrix
\begin{align} \label{eq:V_tilde}
\begin{split}
    \tilde{V} \coloneqq J V.
\end{split}
\end{align}
In the case of Gaussian states, the symplectic picture is equivalent to the density operator description. In the case of other states, it describes a subset of the system's degrees of freedom.

\subsection{Time evolution}
%While a notably simpler object than the density operator, the dynamics of the covariance matrix are not always traceable. This is because most evolution equations for the density operator couple the covariance matrix to higher-order correlations, resulting in a challenging (often infinite) hierarchy of equations not amenable to analytical studies.

It is well-known that the structure-preserving evolution of Gaussian states is governed by Hamiltonians that are second-degree polynomials in mode quadratures:
\begin{align} \label{eq:hamiltonian_quadratic}
    \hat{H} = \frac{1}{2}\hat{\vec{\xi}}^T G \hat{\vec{\xi}},
\end{align}
where $G$ is a $2N\times 2N$, real, symmetric matrix. Similarly, the set of Gaussian states is preserved if the Lindblad operators are assumed to be linear in the mode quadratures, or linear for short:
\begin{align} \label{eq:dissipation_quadratic}
    \hat{L}_j = \vec{c}_j \cdot \hat{\vec{\xi}},\qquad\vec{c}_j\in\mathbb{C}^{2N}.
\end{align}
%The model of evolution based on eqs (\ref{eq:hamiltonian_quadratic}, \ref{eq:dissipation_quadratic}) is especially important from the point of view of engineered dissipation (though it is not limited to it), as, due to technical limitations, contemporary experiments employ at most second-order operators (note that linear Lindblad operators correspond to second-order dissipators). 

However, the type of operations routinely accessible in current experiments on CV systems can be used to manipulate not only Gaussian states, but also their convex combinations. Indeed, resource theories of non-Gaussianity \cite{Gaussianity_resource_Albarelli_2018,Gaussianity_resource_Takagi_2018}, which are built around the set of operations available in contemporary CV experiments, put Gaussian states and their convex combinations on equal footing.

For this reason, in addition to linear Lindblad operators typically assumed in studies of Gaussian systems, we also consider the subclass of Lindblad operators of the form
\begin{align} \label{eq:dissipation_infinite}
    \hat{L}_j = \sqrt{\kappa_j} \hat{U}_j,
\end{align}
where $\hat{U}_j$ are unitary operators, $\kappa_j\geqslant 0$ and $\sum_{j}\kappa_j=1$. Here, it is assumed that the unitaries have quadratic generators, i.e. they can be written as 
\begin{align} \label{eq:unitary_lindbladian_exponential}
    \hat{U}_j = e^{-i \hat{h}_j}
\end{align}
with $\hat{h}_j$ being a Hermitian polynomial of at most second degree in quadrature operators. 

While such Lindblad operators do not preserve the set of Gaussian states, they preserve the set of their convex combinations \cite{quantum_gaussian_evolution_linowski_2022}, making them fully compatible with the resource-theoretic perspective. From the physical point of view,  unitary Lindblad operators constitute a natural model of random noise in the system, with special emphasis on random scattering \cite{unitary_lindbladians_kossakowski_1972,unitary_lindbladians_Kummerer_1987,unitary_lindbladians_Frigerio_1989}.

Computing the time derivative of the covariance matrix and assuming that the system evolves according to the GKLS equation with Hamiltonian (\ref{eq:hamiltonian_quadratic}) and either linear (\ref{eq:dissipation_quadratic}) or unitary (\ref{eq:dissipation_infinite}) Lindblad operators, one can obtain the corresponding equation for the covariance matrix \cite{using_dissipation_1,using_dissipation_2,quantum_gaussian_evolution_linowski_2022}
\begin{align} \label{eq:covariance_evolution_quadratic}
\begin{split}
    \frac{d}{dt}V &= JG V - V GJ  + \mathcal{D}(V). 
\end{split}
\end{align}
Here, the first two terms are responsible for the Hamiltonian evolution, while the symplectic-picture dissipator is equal to
\begin{align} \label{eq:dissipator_covariance_matrix}
\begin{split}
    \mathcal{D}(V) &= \gamma_L \mathcal{D}_L(V) + \gamma_U \mathcal{D}_U(V),\\
    \mathcal{D}_L(V) &=J \im (C^\dag C) V + V \im (C^\dag C) J \\
    & \quad - J \re (C^\dag C) J, \\
    \mathcal{D}_U(V) &=\sum_{j} \kappa_j\left(K_j V K_j^T - V\right),
\end{split}
\end{align}
where $\gamma_L$, $\gamma_U$ are dissipation rates, $\mathcal{D}_L$ comes from linear Lindblad operators and $\mathcal{D}_U$ from their unitary counterparts. Here, $\re (C^\dag C)$ and $\im(C^\dag C)$ denote the real and imaginary parts of the matrix $C^\dag C$, with $C$ defined by the vectors from Eq. (\ref{eq:dissipation_quadratic}) as
\begin{align} \label{eq:C_kj}
\begin{split}
    C_{jk}\coloneqq (\vec{c}_j)_k.
\end{split}
\end{align}
Furthermore, $K_j$ are $2N\times 2N$ symplectic matrices defined by the action of the corresponding unitary Lindbladian on the vector of quadrature operators [all unitaries of the form (\ref{eq:unitary_lindbladian_exponential}) produce such an equation due to the Baker–Campbell–Hausdorff formula]:
\begin{align} \label{eq:K_definition}
\begin{split}
    \hat{U}_j^\dag \hat{\vec{\xi}} \hat{U}_j = K_j \hat{\vec{\xi}}.
\end{split}
\end{align}

\subsection{Stabilizability} The framework of stabilizability can be naturally extended to the covariance matrix evolution. In an analogy to the density operator evolution, here, the Hamiltonian can stabilize the covariance matrix only if the symplectic picture dissipator does not alter the state's symplectic eigenvalues \cite{stabilizability_cv_systems}.

In \cite{stabilizability_cv_systems}, the following necessary conditions for stabilizability of invertible covariance matrices were derived \footnote{In the case of dissipation stemming from linear Lindblad operators, the odd conditions $k=1,3,\ldots$ were shown to be always trivially fulfilled \cite{stabilizing_entanglement_in_two_mode_Gaussian_states}.} for the special case of $\gamma_U=0$, i.e. when $\mathcal{D} = \gamma_L\mathcal{D}_L$:
\begin{align} \label{eq:necessary_conditions_cv}
\begin{split}
    0 = \Tr \left[\tilde{\mathcal{D}}_L(\tilde{V}) \tilde{V}^{k-1} \right] 
    \textnormal{  for all  }k\in\{1,\ldots,2N\},
\end{split}
\end{align}
where
\begin{align} \label{eq:D_L_tilde}
\begin{split}
    \tilde{\mathcal{D}}_L(\tilde{V}) & \coloneqq \{\im (C^\dag C) J, \tilde{V}\} + \re (C^\dag C) J.
\end{split}
\end{align}
As we will now show, eq. (\ref{eq:necessary_conditions_cv}) holds also if $\gamma_U\neq 0$. In other words, we have that
\begin{align} \label{eq:necessary_conditions_cv_general}
\begin{split}
    0 = \Tr \left[\tilde{\mathcal{D}}(\tilde{V}) \tilde{V}^{k-1} \right] 
    \textnormal{  for all  }k\in\{1,\ldots,2N\},
\end{split}
\end{align}
where
\begin{align} \label{eq:D_U_tilde}
\begin{split}
    \tilde{\mathcal{D}}(\tilde{V}) &= 
        \gamma_L \tilde{\mathcal{D}}_L(\tilde{V}) 
        + \gamma_U \tilde{\mathcal{D}}_U(\tilde{V}),\\
    \tilde{\mathcal{D}}_U(\tilde{V}) & \coloneqq 
        \sum_j \kappa_j \left(\tilde{K}_j \tilde{V} \tilde{K}_j^{-1} - \tilde{V}\right)
\end{split}
\end{align}
and $\tilde{K}_j\coloneqq J K_j J^T $. 

To see this, we follow the original derivation \cite{stabilizability_cv_systems}: if the state's symplectic eigenvalues are invariant, then so are any moments of the matrix $\tilde{V}$, since its eigenvalues depend solely on the symplectic eigenvalues of $V$. Written in mathematical notation, for a stabilizable covariance matrix
\begin{align} \label{eq:V_moments_vanishing}
    \frac{d}{dt}\Tr\tilde{V}^k = \Tr \left(\frac{d}{dt}\tilde{V}\right)\tilde{V}^{k-1} 
        =  0.
\end{align}
Note that it is enough to consider $k\leqslant 2N$, as all higher moments necessarily depend on the first $2N$.

The time derivative of $\tilde{V}$ can be computed using eq. (\ref{eq:covariance_evolution_quadratic}). Due to the properties (\ref{eq:J_properties}), we obtain
\begin{align}
\begin{split}
    \frac{d}{dt}\tilde{V} &= [GJ, \tilde{V}] 
        + \gamma_L J\mathcal{D}_L(V) + \gamma_U J\mathcal{D}_U(V).
\end{split}
\end{align}
Substituting this into eq. (\ref{eq:V_moments_vanishing}) we quickly find that the first commutator term vanishes due to the cyclic property of the trace. As for the remaining term, one can easily check by direct calculation that
\begin{align}
\begin{split}
    \Tr \gamma_L J \mathcal{D}_L(V) \tilde{V}^{k-1} &= 
        \Tr \left[\gamma_L\tilde{\mathcal{D}}_L(\tilde{V}) \tilde{V}^{k-1} \right],\\
    \Tr \gamma_U J \mathcal{D}_U(V) \tilde{V}^{k-1} &= 
        \Tr \left[\gamma_U\tilde{\mathcal{D}}_U(\tilde{V}) \tilde{V}^{k-1} \right]
\end{split}
\end{align}
with $\tilde{\mathcal{D}}_L(\tilde{V})$, $\tilde{\mathcal{D}}_U(\tilde{V})$ as in eqs (\ref{eq:D_L_tilde}, \ref{eq:D_U_tilde}). We therefore find that that eq. (\ref{eq:V_moments_vanishing}) is equivalent to
\begin{align}
    \Tr \left(\gamma_L\tilde{\mathcal{D}}_L(\tilde{V}) +
        \gamma_U\tilde{\mathcal{D}}_U(\tilde{V})\right)\tilde{V}^{k-1} =  0,
\end{align}
which concludes our proof due to the first line of eq. (\ref{eq:D_U_tilde}).

Given a stabilizable covariance matrix, one can recover the stabilizing Hamiltonian via  eq. (\ref{eq:hamiltonian_quadratic}) with
\begin{align} \label{eq:stabilizability_hamiltonian_cv}
\begin{split}
    G = \gamma \sum_{\substack{l,l'=0 \\ z_l\neq z_{l'}}}^{d-1}
	    \frac{\vec{w}_l^\dag \tilde{\mathcal{D}}(\tilde{V}) \vec{w}_{l'}}{z_l-z_{l'}} J\sqrt{V}\vec{w}_l\vec{w}_{l'}^\dag\sqrt{V}J^T,
\end{split}
\end{align}
where $\{z_l,\vec{w}_l\}$ is the eigendecomposition of the matrix
\begin{align} \label{eq:mathcal_V}
\begin{split}
    \mathcal{V} \coloneqq \sqrt{V}J\sqrt{V}.
\end{split}
\end{align}
The eigenvalues of $\mathcal{V}$ are connected to the symplectic eigenvalues of $V$ via \cite{stabilizability_cv_systems}
\begin{align} \label{eq:mathcal_V_eigensystem}  
\begin{split}
    z_l = \begin{cases}
        i\nu_l & l=1,\ldots,N, \\
        -i\nu_{N-l} & l=N+1,\ldots,2N.
    \end{cases}
\end{split}
\end{align}

\section{Spectral approach to stabilizability of the covariance matrix}
\label{sec:covariance_matrix_stabilizing}
We are now in a position to extend our spectral approach to stabilizability to the covariance matrix. The main idea is similar to the case of stabilizability of density operators: the covariance matrix can be stabilized by the Hamiltonian only if the dissipator initially leaves its symplectic eigenvalues invariant \cite{stabilizability_cv_systems}. That is, a covariance matrix $V$ may be stabilizable only if its evolution in the absence of the Hamiltonian term
\begin{align} \label{eq:dissipator_evolution_cv}
    \frac{d}{dt}V &= \gamma \mathcal{D}(V), \quad V(0) = V
\end{align}
has a solution $V(t)$ such that
\begin{align} \label{eq:dissipator_solution_cv}
    \mathcal{V}(t) = \sum_{j,j'=1}^{2N} z_j(t) \, \vec{w}_j(t) \vec{w}_j^\dag(t),
        \quad \frac{d z_l(0)}{dt} = 0,
\end{align}
where the eigenvectors $\vec{w}_{l}(t)$ of the matrix $\mathcal{V}(t)$ are orthonormal due to the matrix being asymmetric. Similarly to the case of the density operator, we will now show that the remaining drift of the vectors $\vec{w}_l(t)$ at $t=0$ can be always counteracted by adding an appropriate Hamiltonian term to the equation.

We begin by observing that, due to eq. (\ref{eq:dissipator_solution_cv}), the symplectic eigenvalues of the covariance matrix can be computed as
\begin{equation}
\begin{split}
    z_l(t)\delta_{ll'}=\vec{w}_{l}^\dag(t)\mathcal{V}(t)\vec{w}_{l'}(t).
\end{split}
\end{equation}
Taking the time derivative of both sides at $t=0$, we obtain
\begin{equation}
\begin{split}
    0 = & \, \vec{w}_{l}^\dag\left(\frac{d \sqrt{V}}{dt}J\sqrt{V}
    +\sqrt{V}J\frac{d \sqrt{V}}{dt}\right)\vec{w}_{l'}\\
    & \qquad\qquad\qquad + \frac{d\vec{w}_{l}^\dag}{dt}\mathcal{V}\vec{w}_{l'}
    + \vec{w}_{l}^\dag\mathcal{V}\frac{d\vec{w}_{l'}}{dt}.
\end{split}
\end{equation}
As in the case of stabilizability of the density operator, we skip writing the time dependence explicitly, assuming that all the quantities are evaluated at the initial time. The last two terms vanish due to the orthonormality of the eigenbasis. As for the remaining two terms, we assume $V$ to be invertible and insert
\begin{equation} \label{eq:clever_identity}
\begin{split}
    \mathds{1}_{2N}=\sqrt{V}^{-1}\sqrt{V}, \quad \mathds{1}_{2N}=\sqrt{V}\sqrt{V}^{-1},
\end{split}
\end{equation}
the former in front of $J$ in the first term and the latter after $J$ in the second term. Deploying the eigenrelations of $\mathcal{V}$ results in
\begin{equation}
\begin{split}
    0 = z_l \vec{w}_{l}^\dag\bigg(&\frac{d\sqrt{V}}{dt}\sqrt{V}^{-1}
        +\sqrt{V}^{-1}\frac{d\sqrt{V}}{dt}\bigg)\vec{w}_{l'}.
\end{split}
\end{equation}
Using $V=\sqrt{V}\sqrt{V}$, one can easily see that the above is equivalent to
\begin{equation}
\begin{split}
    0 = z_l & \vec{w}_{l}^\dag\sqrt{V}^{-1}\frac{dV}{dt}\sqrt{V}^{-1}\vec{w}_{l'}.
\end{split}
\end{equation}
Now, the time derivative can be replaced by eq. (\ref{eq:dissipator_evolution_cv}), yielding
\begin{align} \label{eq:theorem_main_cv_almost}
    0=\vec{w}_l^\dag \sqrt{V}^{-1} \mathcal{D}(V) 
        \sqrt{V}^{-1} \vec{w}_{l'}.
\end{align}
Finally, we notice that $\vec{\zeta}_{l'}\coloneqq \sqrt{V}^{-1} \vec{w}_{l'}$ is the eigenvector of the matrix $\tilde{V}$ defined in eq. (\ref{eq:V_tilde}) with eigenvalue $z_{l'}$. Indeed:
\begin{align}
    \tilde{V}\vec{\zeta}_{l'} = J\sqrt{V}\vec{w}_{l'} = \sqrt{V}^{-1}\mathcal{V}\vec{w}_{l'}
        = {z}_{l'} \vec{\zeta}_{l'}.
\end{align}
Thus, eq. (\ref{eq:theorem_main_cv_almost}) is equivalent to
\begin{align} \label{eq:stabilizability_spectral_cv_template}
    0=\vec{\zeta}_l^\dag \mathcal{D}(V) \vec{\zeta}_{l'}
\end{align}
for all $l$, $l'$ such that $z_l=z_{l'}$. It is therefore a necessary condition for stabilizability of the covariance matrix.

However, this condition is also sufficient for for stabilizability. An explicit calculation shows that, provided eq. (\ref{eq:stabilizability_spectral_cv_template}) is fulfilled, eq. (\ref{eq:covariance_evolution_quadratic}) vanishes the input Hamiltonian (\ref{eq:stabilizability_hamiltonian_cv}), i.e. this Hamiltonian stabilizes the covariance matrix: $V(t)=V(0)=V$. Thus, eq. (\ref{eq:stabilizability_spectral_cv_template}) is equivalent to the stabilizability of $V$. 

Once again, we summarize our result in a proposition.

\begin{proposition}[Spectral conditions for stabilizability of the covariance matrix]
\label{th:stabilizability_spectral_cv}
Let $V$ be an \emph{invertible} covariance matrix and $\{z_l,\vec{\zeta}_l\}$ be the eigendecomposition of the matrix $\tilde{V}=JV$. Then, the covariance matrix $V$ is stabilizable with respect to the dissipator $\mathcal{D}$ of the form (\ref{eq:dissipator_covariance_matrix}) if and only if
\begin{align} \label{eq:stabilizability_spectral_cv}
    0=\vec{\zeta}_l^\dag \mathcal{D}(V) \vec{\zeta}_{l'} 
    \textnormal{ for all } l,l' \textnormal{ such that } z_l=z_{l'}.
\end{align}
\end{proposition}

Our previous discussion regarding spectral stabilizability of the density operator can be easily generalized to the covariance matrix. Most importantly, the spectral conditions for the covariance matrix are stronger than the original conditions (\ref{eq:necessary_conditions_cv}), in the sense that they are not only necessary, but also sufficient for stabilizability. Furthermore, they are again only linear in the symplectic eigenvalues, rendering their analysis more tractable.

To illustrate the advantages of the spectral approach for stabilizing covariance matrices, we consider three examples: one for linear, one for unitary Lindblad operators, and one for a mix of the two classes. 

As for the target covariance matrices, we restrict ourselves to covariance matrices in the so-called \emph{standard form}
\begin{align} \label{eq:covariance_matrix_standard_form}
    V_{\textrm{sf}} = 
    \begin{bmatrix}
    a & 0 & c_+ & 0 \\
    0 & a & 0 & c_- \\
    c_+ & 0 & b & 0 \\
    0 & c_- & 0 & b
    \end{bmatrix},
\end{align}
which can be assumed without loss of information whenever one is interested only in the global properties of the state (such as, e.g. entanglement or entropy). Here, $a,b>0$ are proportional to the average number of excitations in the two modes, while $c_{\pm}\in\mathbb{R}$ measure non-local correlations between them \cite{stabilizing_entanglement_in_two_mode_Gaussian_states}.

As discussed above, the spectral approach is most naturally applied to problems in which the target state's eigendecomposition is at least partially emphasized. For this reason, in the case at hand, we restrict ourselves to squeezed thermal states, which are partially parametrized in terms of their symplectic eigenvalues:
\begin{equation} \label{eq:squeezed_parametrization}
\begin{split}
    a &= \nu_1 \cosh^2 r + \nu_2 \sinh^2 r,\\
    b &= \nu_1 \sinh^2 r + \nu_2 \cosh^2 r,\\
    c_\pm &= \pm \frac{\nu_1 + \nu_2}{2}\sinh 2r,
\end{split}
\end{equation}
where $r>0$ is the squeezing strength. Two-mode squeezed thermal states play an important role in quantum metrology, see e.g. \cite{squeezed_metrology,squeezed_review,squeezed_dissipation_experimental_1}.

One can easily calculate the eigensystem of $\tilde{V}_{\textnormal{sf}}$, yielding
\begin{equation} \label{eq:squeezed_eigensystem}
\begin{split}
    z_1 = z_2^* &= i\nu_1, \:\: \vec{\zeta}_1 = \vec{\zeta}_2^* = (-i\coth{r},\coth{r},i,1)^T, \\
    z_3 = z_4^* &= i\nu_2, \:\: \vec{\zeta}_3 = \vec{\zeta}_4^* = (i\tanh{r},\tanh{r},-i,1)^T,
\end{split}
\end{equation}
where we note that the eigenvectors' norm is irrelevant for the spectral conditions (\ref{eq:stabilizability_spectral_cv}). Let us observe that, because $\vec{\zeta}_1 = \vec{\zeta}_2^*$, the corresponding spectral conditions are equivalent, and similarly for $\vec{\zeta}_3 = \vec{\zeta}_4^*$. This means that in the following examples it will be enough to consider only the conditions given by $l=l'=1$ and $l=l'=3$.

\begin{example}[Linear Lindblad operators] \label{ex:stabilizng_entanglement}
Recently, stabilizability was used to investigate the robustness of entangled two-mode Gaussian states against three classes of dissipators based on linear Lindblad operators \cite{stabilizing_entanglement_in_two_mode_Gaussian_states} occurring, e.g. in quantum computation and spectroscopy \cite{Gaussian_states_in_technologies,OPOs_spectroscopy}:
\begin{enumerate}[i.]
    \item Local damping: $\hat{L}_1 \coloneqq \hat{a}_1$ and $\hat{L}_2 \coloneqq \hat{a}_2$;
    \item Damping with global vacuum: $\hat{L} \coloneqq\left(\hat{a}_1+\hat{a}_2\right)$;
    \item Dissipation engineered to preserve two-mode squeezed states with squeezing strength $\alpha$:
        \begin{align}
        \begin{split}
            \hat{L}_1 & \coloneqq 
                \cosh\alpha\, \hat{a}_1 - \sinh\alpha\, \hat{a}_2^\dag, \\
            \hat{L}_2 & \coloneqq
	            \cosh\alpha\, \hat{a}_2 - \sinh\alpha\, \hat{a}_1^\dag.
        \end{split}
        \end{align}
\end{enumerate}
Here, $\hat{a}_k$ is the annihilation operator associated with the $k$-th mode, i.e., $\hat{a}_k\coloneqq \frac{1}{\sqrt{2}}\left(\hat{x}_k+i\hat{p}_k\right)$. Our goal is to use the spectral approach to stabilizability to calculate the set of stabilizable states with respect to each of the above dissipators, which we consider separately so that we can compare with \cite{stabilizing_entanglement_in_two_mode_Gaussian_states}. 

Because all the dissipators are linear, we have $\gamma_U=0$. We only need to calculate the matrices $\im (C^\dag C)$, $\re (C^\dag C)$ entering the dissipator $\mathcal{D}_L$. Recasting the Lindblad operators into the form (\ref{eq:dissipation_quadratic}), computing the matrix $C$ through eq. (\ref{eq:C_kj}) and finally taking the real and imaginary parts of the matrix $C^\dag C$, we obtain:
\begin{enumerate}[i.]
    \item Local damping: $\im C^\dag C = \frac{1}{2} J$ and $\re C^\dag C = \frac{1}{2}\mathds{1}_4$;
    \item Damping with global vacuum:
        \begin{align}
        \begin{split}
            \im C^\dag C = \frac{1}{2}
            \begin{bmatrix}
             J_2 & J_2 \\
             J_2 & J_2
            \end{bmatrix}, \quad
            \re C^\dag C = \frac{1}{2}
            \begin{bmatrix}
             \mathds{1}_2 & \mathds{1}_2 \\
             \mathds{1}_2 & \mathds{1}_2
            \end{bmatrix},
        \end{split}
        \end{align}
        where $J_2$ is as in eq. (\ref{eq:symplectic_form_explicit});
    \item Dissipation engineered to preserve two-mode squeezed states with squeezing strength $\alpha$:
        \begin{align} \label{eq:re3_im3}
        \begin{split}
            \im C^\dag C = \frac{1}{2} J, \: \re C^\dag C &= \frac{1}{2}
            \begin{bmatrix}
            \cosh 2 \alpha \,\mathds{1}_2 & -\sinh 2 \alpha \,\eta_2 \\
            -\sinh 2 \alpha \,\eta_2 & \cosh 2 \alpha \,\mathds{1}_2
            \end{bmatrix},
        \end{split}
        \end{align}
        where
        \begin{align} \label{eq:eta}
            \eta_2 =
            \begin{bmatrix}
            1 & 0 \\
            0 & -1
            \end{bmatrix}.
        \end{align}
\end{enumerate}

Solving the conditions (\ref{eq:stabilizability_spectral_cv}), we find that stabilizable states for the three models are given by
\begin{equation} \label{eq:stabilizing_entanglement_solution}
\begin{split}
    2\nu_1 = 2\nu_2 = \begin{dcases}
        \cosh 2r, & \textnormal{models i. and ii.}, \\
        \cosh 2(r-\alpha), & \textnormal{model iii.},
    \end{dcases}
\end{split}
\end{equation}
which are all physical, as for all of them clearly $\nu_2\geqslant \nu_1 \geqslant 1/2$, as required by eq. (\ref{eq:symplectic_eigenvalues}).

The findings for model i. and iii. coincide with the original results \cite{stabilizing_entanglement_in_two_mode_Gaussian_states}, while the solution for model ii. extends them to the case of squeezed thermal states. 
\end{example}

\begin{example}[Unitary Lindblad operators] \label{ex:IOCM}
For our second example, we consider dissipation stemming from unitary Lindblad operators. To this end, we consider three transformations corresponding to channels often utilized in studies of Gaussianity, entropy and entanglement, among others \cite{example_transformations,De_Palma_2017,squashed_entanglement_De_Palma_2019}:
\begin{enumerate}[i.]
    \item Phase conjugation / transposition channel, with unitary Lindlad operator inducing the following symplectic transformation via eq. (\ref{eq:K_definition}) \cite{example_transformations}:
    \begin{align} \label{eq:K_1}
        K_1 = 
        \begin{bmatrix}
        \sinh\mu \, \eta_2 & \cosh\mu \, \mathds{1}_2 \\
        \cosh\mu \, \mathds{1}_2 & \sinh\mu \, \eta_2
        \end{bmatrix}.
    \end{align}
    \item Beamsplitter / attenuator channel, which correponds to \cite{example_transformations}
    \begin{align} \label{eq:K_2}
        K_2 = 
        \begin{bmatrix}
        \cos\theta \, \mathds{1}_2 & -\sin\theta \, \mathds{1}_2 \\
        \sin\theta \, \mathds{1}_2 & \cos\theta \, \mathds{1}_2
        \end{bmatrix};
    \end{align}
    \item Amplifier channel, which corresponds to \cite{example_transformations}
    \begin{align} \label{eq:K_3}
        K_3 = 
        \begin{bmatrix}
        \cosh\delta \, \mathds{1}_2 & \sinh\delta \, \eta_2 \\
        \sinh\delta \, \eta_2  & \cosh\delta \, \mathds{1}_2
        \end{bmatrix}.
    \end{align}
\end{enumerate}
Here, the parameters $\mu,\delta\in\mathbb{R}$, $\theta\in[0,2\pi)$ quantify the channels' strengths. The more they deviate from the points $\mu=0$, $\theta\in\{0,\pi\}$ and $\delta=0$, which correspond to trivial transformations \footnote{For $\theta\in\{0,\pi\}$ and $\delta=0$, $K_2$ and $K_3$ are proportional to identity. As for $K_1$, the choice $\mu=0$ corresponds to a mere relabeling of the modes: $(\hat{x}_1,\hat{p}_1)\leftrightarrow(\hat{x}_2,\hat{p}_2)$.}, the stronger the channels. Furthermore, $\eta_2$ is as in eq. (\ref{eq:eta}). 

The most general dissipation corresponding to the three channels is given by the bottom line of eq. (\ref{eq:dissipator_covariance_matrix}) with $K_j$ given by eqs (\ref{eq:K_1}-\ref{eq:K_3}). We stress that, unlike in the previous example, we consider the three channels collectively. Assuming the covariance matrix to be in the standard form (\ref{eq:covariance_matrix_standard_form}-\ref{eq:squeezed_parametrization}), we find that the spectral conditions (\ref{eq:stabilizability_spectral_cv}) for stabilizability are fulfilled only in the trivial cases
\begin{align} \label{eq:example_trivial_solution}
\begin{split}
    (\kappa_1=0 \textnormal{ or } \mu=0) \textnormal{ and }
    (\kappa_2=0 \textnormal{ or } \theta\in\{0,\pi\}) \\ \textnormal{ and }
    (\kappa_3=0 \textnormal{ or } \delta=0). 
\end{split}
\end{align}
In other words, there exist no nontrivial stabilizable states with respect to dissipation given by eqs (\ref{eq:K_1}-\ref{eq:K_3}).

To see this, let us consider the case, in which $\kappa_j\neq 0$ for all $j$. The conditions (\ref{eq:stabilizability_spectral_cv}) have the following explicit form:
\begin{align} \label{eq:FG_conditions}
\begin{split}
    0 &= \frac{[Q - 2(\kappa_1+\bar{\kappa}_2)] \nu_1 
        + [Q + 2(\kappa_1+\bar{\kappa}_2)] \nu_2}{\sinh^2 r}, \\
    0 &= \frac{[Q + 2(\kappa_1+\bar{\kappa}_2)] \nu_1 
        + [Q - 2(\kappa_1+\bar{\kappa}_2)] \nu_2}{\cosh^2 r},
\end{split}
\end{align}
where $\bar{\kappa}_2 \equiv \kappa_2 \sin^2\theta$ and
\begin{align} \label{eq:Q}
\begin{split}
    Q \equiv \kappa_1\left(\cosh 2\mu-1\right) 
        + \bar{\kappa}_2\left(\cosh 4r - 1\right)
        + \kappa_3\left(\cosh 2\nu-1\right).   
\end{split}
\end{align}
Clearly, eq. (\ref{eq:FG_conditions}) is fulfilled only if the two numerators vanish. If so, then their difference must also vanish, which yields, after simplication,
\begin{align}
\begin{split}
    0 = (\nu_1-\nu_2)\left(\kappa_1 + \bar{\kappa}_2\right).
\end{split}
\end{align}
Under our assumption that $\kappa_j\neq 0$, the only solution to this equation is $\nu_1=\nu_2$. Then, both of the two equations (\ref{eq:FG_conditions}) reduce to
\begin{align}
\begin{split}
    0 = Q,
\end{split}
\end{align}
which, as we can see clearly from eq. (\ref{eq:Q}), has no non-trivial solutions. This finishes the proof for the case $\kappa_j\neq 0$ for all $j$. The remaining special cases can be treated analogously.

To derive our result, we heavily used the fact that the spectral conditions are linear in the covariance matrix' symplectic eigenvalues. It allowed us to obtain an easily solvable equation after substracting the two original conditions from each other, and then get a condition independent from the symplectic eigenvalues upon setting $\nu_1=\nu_2$. Similar operations are not applicable using the corresponding geometric conditions (\ref{eq:necessary_conditions_cv}), which in this case consist of two polynomial equations of second and fourth order in these eigenvalues. This implies a much higher computational complexity than in the case of the linear spectral conditions, and consequently we were unable to analytically rederive the result (\ref{eq:example_trivial_solution}) using the geometric approach. 
\end{example}

\begin{example}[Mixed Lindblad operators]
As a final example, let us consider a mixed dissipator. Specifically, we assume that part of the dissipation is engineered to preserve two-mode squeezed states with squeezing strength $\alpha$, given by a linear Lindblad operator as in model iii from Example \ref{ex:stabilizng_entanglement}, while the system is disturbed by additional amplification given by a unitary Lindblad operator, as in model iii from Example \ref{ex:IOCM}. In other words, we assume the dissipator (\ref{eq:dissipator_covariance_matrix}) with $\re C^\dag C$, $\im C^\dag C$ as in eq. (\ref{eq:re3_im3}), $\kappa_j=\delta_{j3}$ and $K_3$ as in eq. (\ref{eq:K_3}).

The spectral conditions (\ref{eq:stabilizability_spectral_cv}) read
\begin{align} \label{eq:ex_6_conditions}
\begin{split}
    0 &= \frac{\gamma_L [\cosh(2r-2\alpha)-2\nu_1] 
    + \gamma_U(\cosh 2\delta-1)(\nu_1+\nu_2)}
        {\sinh^2 r}, \\
    0 &= \frac{\gamma_L [\cosh(2r-2\alpha)-2\nu_2] 
    + \gamma_U(\cosh 2\delta-1)(\nu_1+\nu_2)}
        {\cosh^2 r}. \\
\end{split}
\end{align}
As in the previous example, by substracting the numerators from each other we find that $\nu_1 = \nu_2 \equiv \nu$ is necessary for stabilizability. Solving the conditions with this input, we immediately find the ultimate solution
\begin{align}
\begin{split}
    \nu = \frac{\gamma_L\cosh(2r-2\alpha)}
        {2\left[\gamma_L-\gamma_U(\cosh 2\delta - 1)\right]},
\end{split}
\end{align}
which corresponds to a valid covariance matrix (we must have $\nu\geqslant 1/2$) as long as
\begin{align} \label{eq:gamma_L_gamma_U_relation}
\begin{split}
    \gamma_L \geqslant \gamma_U(\cosh 2\delta - 1).
\end{split}
\end{align}
\end{example}
To see what the inclusion of the unitary Lindblad operator changes compared to the original solution, given by the bottom line of eq. (\ref{eq:stabilizing_entanglement_solution}), we compare the amount of entanglement corresponding to the two solutions. To measure entanglement, we deploy the logarithmic negativity, which for two-mode Gaussian states reads \cite{two-mode_gaussian_etc_proper_norm}
\begin{equation} \label{eq:log_neg}
\begin{split}
    E_\mathcal{N} \coloneqq \max 
    \left\{0,-\ln\left[2\tilde{\nu}_-\right]\right\},
\end{split}
\end{equation}
where
\begin{equation} \label{nu_+-tilde}
\begin{split}
\tilde{\nu}_-(V) = 
	\sqrt{\frac{1}{2}\left(\tilde{\Delta}(V)-\sqrt{\tilde{\Delta}^2(V)-4\det V}\right)}
\end{split}
\end{equation}
and $\tilde{\Delta}(V) \coloneqq a^2 + b^2 - 2 c_+ c_-$ in the notation from Eq. (\ref{eq:covariance_matrix_standard_form}).

We find that
\begin{equation} \label{eq:E_N_two-mode}
\begin{split}
    E_\mathcal{N}=
        & - \ln \left[e^{-2r}\cosh 2(r-\alpha)\right] \\
        & - \ln \frac{\gamma_L}{\gamma_L-\gamma_U(\cosh 2\delta - 1)},
\end{split}
\end{equation}
where the second term vanishes for $\gamma_U = 0$, reproducing the original result from \cite{stabilizing_entanglement_in_two_mode_Gaussian_states}. Because of the condition (\ref{eq:gamma_L_gamma_U_relation}), the second term is never positive and thus contributes negatively to the amount of entanglement in the state. This is an expected result, as amplification given by the amplifier channel is typically interpreted as a random process, which generically does not result in an increase of a useful resource such as entanglement.

We note that, similarly to the previous example, we were not able to solve this problem analytically using the geometric conditions due to their computational complexity.

\section{Concluding remarks}
\label{sec:conclusion}
The concept of stabilizability serves to fathom the prospects and limits of coherent control for counteracting the detrimental effects of dissipation in quantum systems. We developed a spectral approach to stabilizability, where the stabilizability conditions manifestly refer to the eigenstates of the state to be stabilized. These spectral conditions complement the previously formulated geometric stabilizability conditions, extending the scope of applicability of the stabilization framework both from a conceptual and a practical perspective, and both in finite-dimensional Hilbert spaces and in Gaussian quantum systems. We presented several examples that exposed the advantages of the spectral over the geometric approach in these cases.

Remarkably, the spectral conditions make it possible to directly address the stabilizability of desired target states as eigenstates, which arguably represents the most informative way to assess how a target state's functionality can be uphold in a dissipation-induced mixed state. We demonstrated this, for instance, with generalized GHZ and W states, where we could identify scaling laws governing their stabilizability for general spin numbers $N$. As we argued, a similar analysis using the geometric conditions and based on maximizing the fidelity with the target state is not possible, since the target state is in general not an eigenstate of the fidelity-optimal mixed state. More generically, the spectral conditions allow us to discuss the stabilizability of dominant eigenstates, the relevance of which has recently been identified, for instance, in the context of quantum state tomography \cite{Melkani2020eigenstate} and quantum error mitigation \cite{Koczor2021exponential,Huggins2021virtual}.

As a final remark, let us stress that the advantage of the spectral approach over the original one in the examples considered by us stems from the fact that we were interested in classes of states whose eigendecomposition was at least partially known. It would be interesting to see whether one can find a set of stabilizability conditions that would have the best of the spectral and geometric approaches and be relatively easy to solve in general.

\acknowledgements 
Tomasz Linowski and {\L}ukasz Rudnicki acknowledge support by the Foundation for Polish Science (International Research Agenda Programme project, International Centre for Theory of Quantum Technologies, Grant No. 2018/MAB/5, cofinanced by the European Union within the Smart Growth Operational Programme).

\bibliography{report}{}

\begin{thebibliography}{10}
\providecommand{\url}[1]{\texttt{#1}}
\providecommand{\urlprefix}{URL }
\providecommand{\eprint}[2][]{\url{#2}}

\bibitem{decoherence_theory_Zurek_1991}
W.~H. Zurek, \emph{Decoherence and the transition from quantum to classical},
  Phys. Today \textbf{44}, 36 (1991).

\bibitem{decoherence_theory_Schlosshauer_2005}
M.~Schlosshauer, \emph{Decoherence, the measurement problem, and
  interpretations of quantum mechanics}, Rev. Mod. Phys. \textbf{76}, 1267
  (2005).

\bibitem{decoherence_quantum_computing_Preskill_1998}
J.~Preskill, \emph{Quantum computing: pro and con}, Proc. R. Soc. Lond. A
  \textbf{454}, 469 (1998).

\bibitem{decoherence_quantum_computing_Lidar_1998}
D.~A. Lidar, I.~L. Chuang, K.~B. Whaley, \emph{Decoherence-free subspaces for
  quantum computation}, Phys. Rev. Lett. \textbf{81}, 2594 (1998).

\bibitem{decoherence_Lidar_2014}
D.~A. Lidar, \emph{Review of Decoherence-Free Subspaces, Noiseless Subsystems,
  and Dynamical Decoupling}, 295--354, John Wiley and Sons, Ltd (2014).

\bibitem{GKS_original}
V.~Gorini, A.~Kossakowski, E.~C.~G. Sudarshan, \emph{Completely positive
  dynamical semigroups of {N}‐level systems}, J. Math. Phys. \textbf{17}, 821
  (1976).

\bibitem{lindblad_original}
G.~Lindblad, \emph{On the generators of quantum dynamical semigroups}, Comm.
  Math. Phys. \textbf{48}, 119 (1976).

\bibitem{using_dissipation_1}
S.~Mancini, H.~M. Wiseman, \emph{Optimal control of entanglement via quantum
  feedback}, Phys. Rev. A \textbf{75}, 012330 (2007).

\bibitem{Kraus2008preparation}
B.~Kraus, H.~P. B\"uchler, S.~Diehl, A.~Kantian, A.~Micheli, et~al.,
  \emph{Preparation of entangled states by quantum {M}arkov processes}, Phys.
  Rev. A \textbf{78}, 042307 (2008).

\bibitem{using_dissipation_2}
K.~Koga, N.~Yamamoto, \emph{Dissipation-induced pure {Gaussian} state}, Phys.
  Rev. A \textbf{85}, 022103 (2012).

\bibitem{Lyapunov_stationarity}
F.~Nicacio, M.~Paternostro, A.~Ferraro, \emph{Determining stationary-state
  quantum properties directly from system-environment interactions}, Phys. Rev.
  A \textbf{94}, 052129 (2016).

\bibitem{using_hamiltonian_1}
S.~Sauer, C.~Gneiting, A.~Buchleitner, \emph{Stabilizing entanglement in the
  presence of local decay processes}, Phys. Rev. A \textbf{89}, 022327 (2014).

\bibitem{using_hamiltonian_2}
M.~Mamaev, L.~C.~G. Govia, A.~A. Clerk, \emph{Dissipative stabilization of
  entangled cat states using a driven {B}ose-{H}ubbard dimer}, Quantum
  \textbf{2}, 58 (2018).

\bibitem{stabilizability_geometric}
S.~Sauer, C.~Gneiting, A.~Buchleitner, \emph{Optimal coherent control to
  counteract dissipation}, Phys. Rev. Lett. \textbf{111}, 030405 (2013).

\bibitem{stabilizability_cv_systems}
{\L}.~Rudnicki, C.~Gneiting, \emph{Stabilizable {Gaussian} states}, Phys. Rev.
  A \textbf{98}, 032120 (2018).

\bibitem{stabilizing_entanglement_in_two_mode_Gaussian_states}
T.~Linowski, C.~Gneiting, {\L}.~Rudnicki, \emph{Stabilizing entanglement in
  two-mode {Gaussian} states}, Phys. Rev. A \textbf{102}, 042405 (2020).

\bibitem{quantum_gaussian_evolution_linowski_2022}
T.~Linowski, A.~Teretenkov, {\L}.~Rudnicki, \emph{Dissipative evolution of
  quantum {Gaussian} states}, Phys. Rev. A \textbf{106}, 052206 (2022).

\bibitem{open_systems_book_Rivas_2012}
A.~Rivas, S.~F. Huelga, \emph{{Open Quantum Systems}}, Springer, 1st edition
  (2012).

\bibitem{quantum_information_book}
M.~A. Nielsen, I.~L. Chuang, \emph{Quantum Computation and Quantum Information:
  10th Anniversary Edition}, Cambridge University Press, 10th edition (2011).

\bibitem{quantum_algorithms}
A.~Montanaro, \emph{Quantum algorithms: {An} overview}, npj Quantum Inf.
  \textbf{2}, 15023 (2016).

\bibitem{Note1}
Necessity can be easily shown by multiplying eq. (\ref
  {eq:startionarity_condition}) by $\protect \hat {X}$ and taking the trace.
  Then, the requirement $[\protect \hat {X}, \protect \hat {\rho }]=0$ makes
  the Hamiltonian part vanish. Sufficiency follows by direct inspection.

\bibitem{three_qubits_entanglement_types}
W.~D\"ur, G.~Vidal, J.~I. Cirac, \emph{Three qubits can be entangled in two
  inequivalent ways}, Phys. Rev. A \textbf{62}, 062314 (2000).

\bibitem{entanglement_resource_theory_unique_maximal_Contreras_2019}
P.~Contreras-Tejada, C.~Palazuelos, J.~I. de~Vicente, \emph{Resource theory of
  entanglement with a unique multipartite maximally entangled state}, Phys.
  Rev. Lett. \textbf{122}, 120503 (2019).

\bibitem{generalized_GHZ_state_computing_2019}
D.~Cruz, R.~Fournier, F.~Gremion, A.~Jeannerot, K.~Komagata, et~al.,
  \emph{Efficient quantum algorithms for {GHZ and W} states, and implementation
  on the {IBM} quantum computer}, Adv. Quantum Technol. \textbf{2}, 1900015
  (2019).

\bibitem{Note2}
Let $\protect \hat {A}$ have the eigendecomposition $\protect \hat {A}=\Sigma
  _j A_j {|{A_j}\rangle }{\langle {A_j}|}$. Then $\protect \Tr \protect \hat
  {\sigma } \protect \hat {A} = \Sigma _{j} A_j \protect \Tr \protect \hat
  {\sigma } {|{A_j}\rangle }{\langle {A_j}|} \leqslant (\protect \qopname
  \relax m{max}_l A_l) \protect \Tr \protect \hat {\sigma } \Sigma _{j}
  {|{A_j}\rangle }{\langle {A_j}|} = (\protect \qopname \relax m{max}_l A_l)
  \protect \Tr \protect \hat {\sigma } = \protect \qopname \relax m{max}_l
  A_l$.

\bibitem{Bell}
J.~S. Bell, \emph{On the {Einstein} {Podolsky} {Rosen} paradox}, Phys. Phys.
  Fiz. \textbf{1}, 195 (1964).

\bibitem{quantum_teleportation}
C.~H. {Bennett}, G.~{Brassard}, C.~{Crepeau}, R.~{Jozsa}, A.~{Peres}, et~al.,
  \emph{{Teleporting an unknown quantum state via dual classical and
  {Einstein}-{Podolsky}-{Rosen} channels}}, Phys. Rev. Lett. \textbf{70}, 1895
  (1993).

\bibitem{Note3}
In fact, based on our findings we conjecture more generally that for all $N$
  the eigenvalues of $\protect \hat {\zeta }$ are: $2/N$ with no degeneracy,
  $(N - 2)/(N (N - 1))$ with degeneracy $N-1$ and zero with degeneracy $2^N-N$.

\bibitem{W_GHZ_comparison_robustness_Carvahlo_2004}
A.~R.~R. Carvalho, F.~Mintert, A.~Buchleitner, \emph{Decoherence and
  multipartite entanglement}, Phys. Rev. Lett. \textbf{93}, 230501 (2004).

\bibitem{Ivan_2012}
J.~S. Ivan, N.~Mukunda, R.~Simon, \emph{{Moments of non-Gaussian Wigner
  distributions and a generalized uncertainty principle: I. The single-mode
  case}}, J. Phys. A: Math. Theor. \textbf{45}, 195305 (2012).

\bibitem{two-mode_gaussian_etc_proper_norm}
G.~Adesso, A.~Serafini, F.~Illuminati, \emph{Determination of continuous
  variable entanglement by purity measurements}, Phys. Rev. Lett. \textbf{92},
  087901 (2004).

\bibitem{cv_systems_gaussian_states}
G.~Adesso, S.~Ragy, A.~R. Lee, \emph{Continuous variable quantum information:
  {Gaussian} states and beyond}, Open Syst. Inf. Dyn. \textbf{21}, 1440001
  (2014).

\bibitem{gaussian_optics}
L.~Mandel, E.~Wolf, \emph{Optical Coherence and Quantum Optics}, Cambridge
  University Press (1995).

\bibitem{Note4}
More precisely, for an $N$-mode system, there are $2N$ first moments and $4N^2$
  second moments. Taking into account the fact that some of the second moments
  are co-dependent (due to the canonical commutation relations) yields a total
  of $N(2N+3)$ independent parameters.

\bibitem{Note5}
Some define symplectic matrices by an alternative relation $K^T J K = J$.
  However, both definitions are completely equivalent, since when $K$ is
  symplectic, so is $K^T$. Indeed, assume, e.g., $K J K^T = J$ holds. Then $K^T
  = J K^{-1} J^{-1}$, which substituted into $K^T J K$ yields $J$. Proof in the
  other direction is analogous.

\bibitem{Gaussianity_resource_Albarelli_2018}
F.~Albarelli, M.~G. Genoni, M.~G.~A. Paris, A.~Ferraro, \emph{Resource theory
  of quantum {non-Gaussianity and Wigner} negativity}, Phys. Rev. A
  \textbf{98}, 052350 (2018).

\bibitem{Gaussianity_resource_Takagi_2018}
R.~Takagi, Q.~Zhuang, \emph{Convex resource theory of {non-Gaussianity}}, Phys.
  Rev. A \textbf{97}, 062337 (2018).

\bibitem{unitary_lindbladians_kossakowski_1972}
A.~Kossakowski, \emph{On quantum statistical mechanics of non-{Hamiltonian}
  systems}, Rep. Math. Phys. \textbf{3}, 247 (1972).

\bibitem{unitary_lindbladians_Kummerer_1987}
B.~K\H{u}mmerer, H.~Maassen, \emph{The essentially commutative dilations of
  dynamical semigroups on {$M_n$}}, Comm. Math. Phys. \textbf{109}, 1 (1987).

\bibitem{unitary_lindbladians_Frigerio_1989}
A.~Frigerio, H.~Maassen, \emph{Quantum {Poisson} processes and dilations of
  dynamical semigroups}, Probab. Theory Relat. Fields \textbf{83}, 489 (1989).

\bibitem{Note6}
In the case of dissipation stemming from linear Lindblad operators, the odd
  conditions $k=1,3,\protect \ldots $ were shown to be always trivially
  fulfilled \cite {stabilizing_entanglement_in_two_mode_Gaussian_states}.

\bibitem{squeezed_metrology}
C.~M. Caves, \emph{Quantum-mechanical noise in an interferometer}, Phys. Rev. D
  \textbf{23}, 1693 (1981).

\bibitem{squeezed_review}
R.~Schnabel, \emph{Squeezed states of light and their applications in laser
  interferometers}, Phys. Rep. \textbf{684}, 1  (2017).

\bibitem{squeezed_dissipation_experimental_1}
C.~A. Muschik, E.~S. Polzik, J.~I. Cirac, \emph{Dissipatively driven
  entanglement of two macroscopic atomic ensembles}, Phys. Rev. A \textbf{83},
  052312 (2011).

\bibitem{Gaussian_states_in_technologies}
S.~Tserkis, T.~C. Ralph, \emph{Quantifying entanglement in two-mode {G}aussian
  states}, Phys. Rev. A \textbf{96}, 062338 (2017).

\bibitem{OPOs_spectroscopy}
M.~Vaidyanathan, R.~C. Eckardt, V.~Dominic, L.~E. Myers, T.~P. Grayson,
  \emph{Cascaded optical parametric oscillations}, Opt. Express \textbf{1}, 49
  (1997).

\bibitem{example_transformations}
J.~S. Ivan, K.~K. Sabapathy, R.~Simon, \emph{Operator-sum representation for
  bosonic {Gaussian} channels}, Phys. Rev. A \textbf{84}, 042311 (2011).

\bibitem{De_Palma_2017}
G.~De~Palma, \emph{The {Wehrl} entropy has {Gaussian} optimizers}, Lett. Math.
  Phys. \textbf{108}, 97–116 (2017).

\bibitem{squashed_entanglement_De_Palma_2019}
G.~De~Palma, \emph{The squashed entanglement of the noiseless quantum
  {Gaussian} attenuator and amplifier}, J. Math. Phys. \textbf{60}, 112201
  (2019).

\bibitem{Note7}
For $\theta \in \{0,\pi \}$ and $\delta =0$, $K_2$ and $K_3$ are proportional
  to identity. As for $K_1$, the choice $\mu =0$ corresponds to a mere
  relabeling of the modes: $(\protect \hat {x}_1,\protect \hat
  {p}_1)\leftrightarrow (\protect \hat {x}_2,\protect \hat {p}_2)$.

\bibitem{Melkani2020eigenstate}
A.~Melkani, C.~Gneiting, F.~Nori, \emph{Eigenstate extraction with
  neural-network tomography}, Phys. Rev. A \textbf{102}, 022412 (2020).

\bibitem{Koczor2021exponential}
B.~Koczor, \emph{Exponential error suppression for near-term quantum devices},
  Phys. Rev. X \textbf{11}, 031057 (2021).

\bibitem{Huggins2021virtual}
W.~J. Huggins, S.~McArdle, T.~E. O'Brien, J.~Lee, N.~C. Rubin, et~al.,
  \emph{Virtual distillation for quantum error mitigation}, Phys. Rev. X
  \textbf{11}, 041036 (2021).

\end{thebibliography}
%\addcontentsline{toc}{chapter}{Bibliography}
\bibliographystyle{obib}

\end{document}